\documentclass[journal]{IEEEtran}

\usepackage{amsmath}
\usepackage{amssymb}
\usepackage{graphicx}
\usepackage{booktabs}
\usepackage{multirow}
\usepackage{array}
\usepackage{algorithm}
\usepackage{algorithmic}
\usepackage{url}
\usepackage{cite}
\usepackage{stfloats}
\usepackage{subcaption}

\newcommand{\Hd}{\bar{H}}
\newcommand{\BER}{\mathrm{BER}}
\newcommand{\SE}{\mathrm{SE}}

\newcommand{\given}{\,|\,}

\begin{document}

\title{PRB-WM: Physics-Regularized Beta World Models for Predictive Link Adaptation on Real Time-Varying Underwater Acoustic Channels}

\author{Shengjie Zhong,~\IEEEmembership{Graduate Student Member,~IEEE,}
        Zhongliang Zhao,
        Jingxuan Chen,
        Dianrun Huang,
        Zichen Zhao,
        Yang Liu,~\IEEEmembership{Senior Member, ~IEEE}
        Dapeng O. Wu,~\IEEEmembership{Fellow,~IEEE},
        Tony Q. S. Quek,~\IEEEmembership{Fellow,~IEEE}
\thanks{S. Zhong is with the School of Electronic and Information Engineering, Beihang University, Beijing 100191, China (e-mail: zhongshengjie@buaa.edu.cn).}
\thanks{Z. Zhao is with the School of Electronic and Information Engineering, Beihang University, Beijing, with the Department of Strategic and Advanced Interdisciplinary Research, Peng Cheng Laboratory, with the Hangzhou International Innovation Institute, Beihang University, Hangzhou, with the MIIT Key Lab. of Aerospace Mobile Communication, Beijing, and also with the State Key Lab. of CNS/ATM, Beijing (e-mail: zhaozl@buaa.edu.cn).}
\thanks{J. Chen is with the Department of Computer Science, City University of Hong Kong, Hong Kong, SAR, China, and also with the Hong Kong Generative AI Research and Development Center, Hong Kong, SAR, China (e-mail: chenjingxuan@buaa.edu.cn).}
\thanks{D. Huang is with the School of Electronic and Information Engineering, Beihang University, Beijing 100191, China (e-mail: zhongshengjie@buaa.edu.cn).}
\thanks{Z. Zhao is with the College of Underwater Acoustic Engineering, Harbin Engineering University, Harbin 150000, China (e-mail: 0424zzc@hrbeu.edu.cn).}
\thanks{Y. Liu is with the College of Computing \& Data Science, Nanyang Technological University, Singapore (e-mail: yangliu@ntu.edu.sg).}
\thanks{D. O. Wu is with the Department of Computer Science, The City University of Hong Kong, Hong Kong, SAR, China (e-mail: dpwu@ieee.org).}
\thanks{T. Q. S. Quek is with the Information Systems Technology and Design Pillar, Singapore University of Technology and Design, Singapore 487372 (e-mail: tonyquek@sutd.edu.sg).}
}

\markboth{IEEE trans, Vol.~XX, No.~XX, XXX~20XX}%
{XXX: PRB-WM for Predictive Link Adaptation on Real Time-Varying UWA Channels}
\IEEEspecialpapernotice{This work has been submitted to the IEEE for possible publication. Copyright may be transferred without notice, after which this version may no longer be accessible.}

\maketitle

\begin{abstract}
Adaptive modulation and coding (AMC) improves the spectral efficiency of underwater acoustic (UWA) links, but its feedback is inevitably stale: sound propagates slowly, and by the time channel state information (CSI) arrives, the fast time-varying multipath channel has moved on. Predictive link adaptation therefore needs more than a point forecast: it needs a distribution over each candidate action's consequences. We propose PRB-WM, a physics-regularized Beta world model for predictive per-subcarrier link adaptation on real time-varying UWA channels. A persistence-anchored stochastic latent-space dynamics model learns how the channel evolves; a counterfactual outcome head evaluates, for every candidate modulation and coding scheme (MCS), a Beta--binomial distribution of bit errors on shared imagined futures; and a chance-constrained receding-horizon model predictive controller maximizes expected spectral efficiency subject to a bounded probability of violating the bit error rate (BER) target. To the best of our knowledge, PRB-WM is the first world model for UWA communications. In closed-loop replay of six measured sea-trial links from the public UWA Channel Library against a 24-method pool spanning the full methodological spectrum, PRB-WM attains a mean goodput $87\%$ above the strongest baseline, ranks first on all six time-varying links at the longest lookahead, and is the only method whose realized frame risk decreases systematically with lookahead, meeting the BER-violation contract where no baseline does. Ablations confirm that rollout sampling dispersion is essential to the chance constraint and that the physics anchor is necessary for trainability.
\end{abstract}


\section{Introduction}
\IEEEPARstart{U}{nderwater} acoustic (UWA) channels are widely regarded as the most difficult wireless medium in use. Sound propagates five orders of magnitude slower than radio, multipath delay spreads are long, and the channel varies quickly in both time and frequency under surface motion, internal waves, and platform drift \cite{stojanovic2009underwater,li2025underwater}. Over the past two decades, multicarrier modulation \cite{li2008multicarrier} and machine-learning-based physical-layer design \cite{huang2022machine} for such channels have matured considerably, and adaptive modulation and coding (AMC) remains the standard mechanism for closing the gap between achieved and available spectral efficiency \cite{cooper2026state}. AMC, however, is only as good as the channel knowledge it is fed. In UWA links the feedback loop is dominated by propagation delay, so the channel state information (CSI) that reaches the transmitter describes a channel that no longer exists. Link adaptation under such latency is therefore inherently a \emph{prediction} problem: the transmitter must decide the modulation and coding scheme (MCS) of future frames from a forecast of how the channel will evolve \cite{huang2020adaptive}.

\begin{figure*}[!t]
\centering
\includegraphics[width=0.85\textwidth]{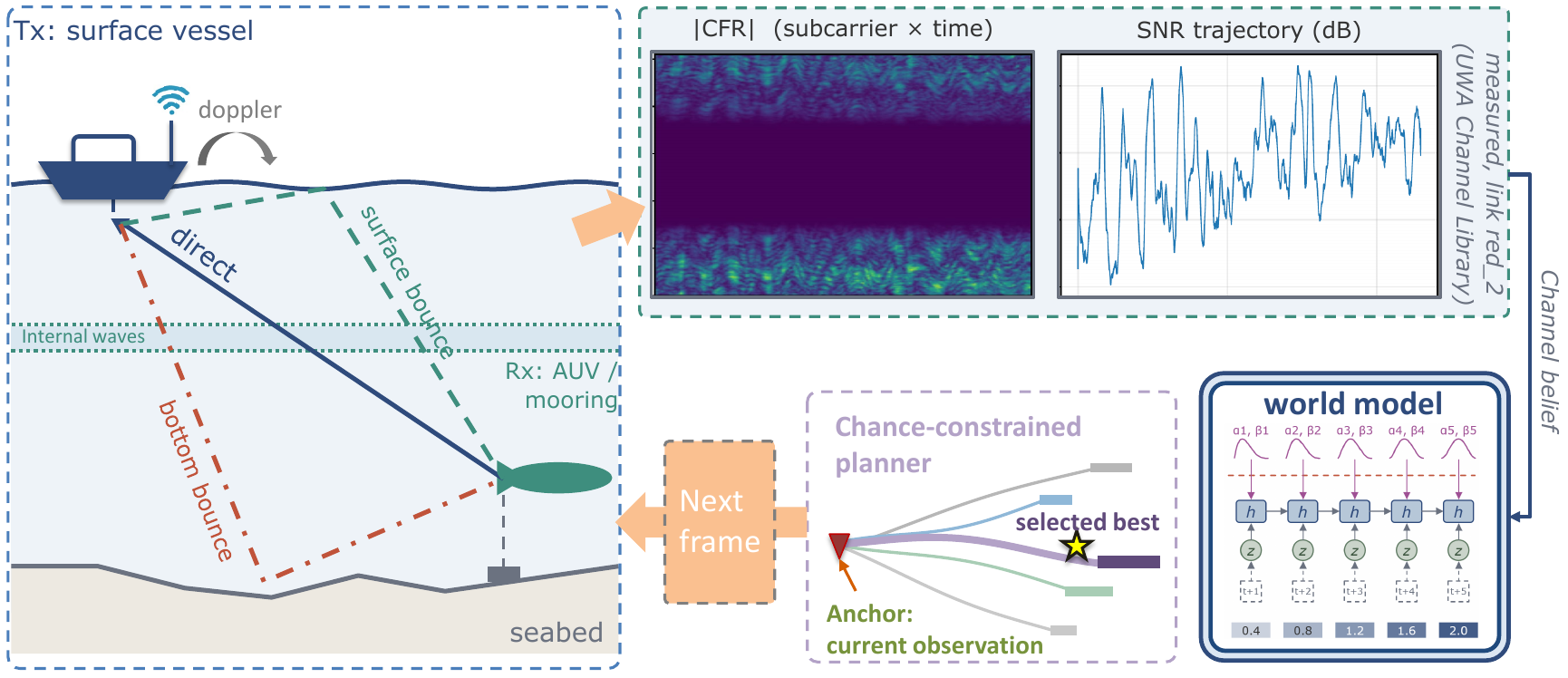}
\caption{Scenario and decision loop. Left: a single-hop UWA orthogonal frequency-division multiplexing (OFDM) link between a surface-vessel transmitter and an autonomous underwater vehicle (AUV)/mooring receiver; direct, surface-bounce, and bottom-bounce paths, internal waves, and platform Doppler make the channel fast time-varying, and the long acoustic propagation delay makes the feedback CSI stale at decision time. Top right: measured channel frequency response (CFR) magnitude and per-frame signal-to-noise ratio (SNR) trajectory of the sea-trial link red\_2 from the UWA Channel Library, illustrating the fast time-frequency variation that motivates stochastic prediction. Bottom right: at the start of every epoch, the world model rolls out a distribution of futures anchored at the current observation, the chance-constrained planner scores every candidate MCS on these imagined futures, and the selected per-subcarrier allocation is transmitted on the next frame.}
\label{fig:scenario}
\end{figure*}

Three families of solutions have made substantial progress toward this goal, and each has a characteristic limitation. Stochastic predictive AMC---Markov chains, hidden Markov models (HMMs), and Kalman filters over SNR-binned channel states---has demonstrated tangible gains on measured UWA links and remains the strongest classical approach \cite{busacca2024predictive,busacca2024adaptive}; its coarse first-order state, however, cannot represent high-dimensional frequency-selective CFRs, and action consequences are modeled weakly. Direct deep multi-step forecasters, such as patch-based Transformer CSI predictors \cite{tian2026patchcsi,jiang2022accurate}, achieve strong point-prediction accuracy over long horizons; but they output deterministic predictions in one shot, carrying no notion of environment dynamics, no predictive distribution, and no model of what an action will do. Online bandit adaptation \cite{busacca2025amuse,busacca2026bandits} learns effective action policies with minimal assumptions and little feedback; it bypasses channel dynamics entirely, however, and cannot answer how the channel will evolve over the next several frames, while variants that select a single link-wide MCS fail structurally on deep-fading frequency-selective channels. \textbf{The common gap is the absence of a model that treats the channel as a learnable stochastic dynamical system and exposes the \emph{distribution} of each candidate action's consequence over a future horizon.}

The measured sea-trial links released with the UWA Channel Library \cite{li2026channel} (Fig.~\ref{fig:scenario}) show exactly the regime in which such a model is needed. The channels are strongly non-stationary: per-frame SNR drifts and fluctuates by several dB, and the CFR magnitude pattern reorganizes within tens of OFDM symbols. Yet the fluctuations are not white: once a simple physics prior is factored out, the residual exhibits short-range correlation that a stochastic dynamics model can exploit. Our closed-loop study of Section~\ref{sec:exp} adds a second, less obvious observation: the forecaster with the best bare normalized mean-square error (NMSE)---a patch-based Transformer---collapses in closed loop, whereas a model with mid-pack NMSE but calibrated stochastic rollouts delivers the highest goodput and the only frame risk that decreases with lookahead. Predictive accuracy and decision value are decoupled; the bottleneck is not a more accurate point forecast but an uncertainty-aware, counterfactually evaluable model of action consequences.

World models offer precisely this capability. Originating in model-based reinforcement learning, where an agent learns a compact latent dynamics model of its environment and plans by imagination \cite{ha2018world,hafner2020dream,hafner2025mastering}, world models have recently crossed into wireless communications: a world model of mobile network traffic supports counterfactual extrapolation under network-parameter adjustments \cite{qi2026mobiwm}, telecom world models have been proposed as a unifying architecture for action-conditioned, uncertainty-aware prediction of network behavior \cite{zou2026telecom}, and recent surveys position world models as the predictive core of digital-twin-driven network intelligence \cite{zheng2026digital}. The UWA link-adaptation setting is, if anything, a better match for the paradigm: the dynamics are governed by ocean physics rather than by the agent's own actions, feedback is severely delayed, and every decision must be justified per subcarrier under an explicit reliability constraint. Yet, to the best of our knowledge, no world model has been developed for UWA channels.

\begin{figure*}[!t]
\centering
\includegraphics[width=0.9\textwidth]{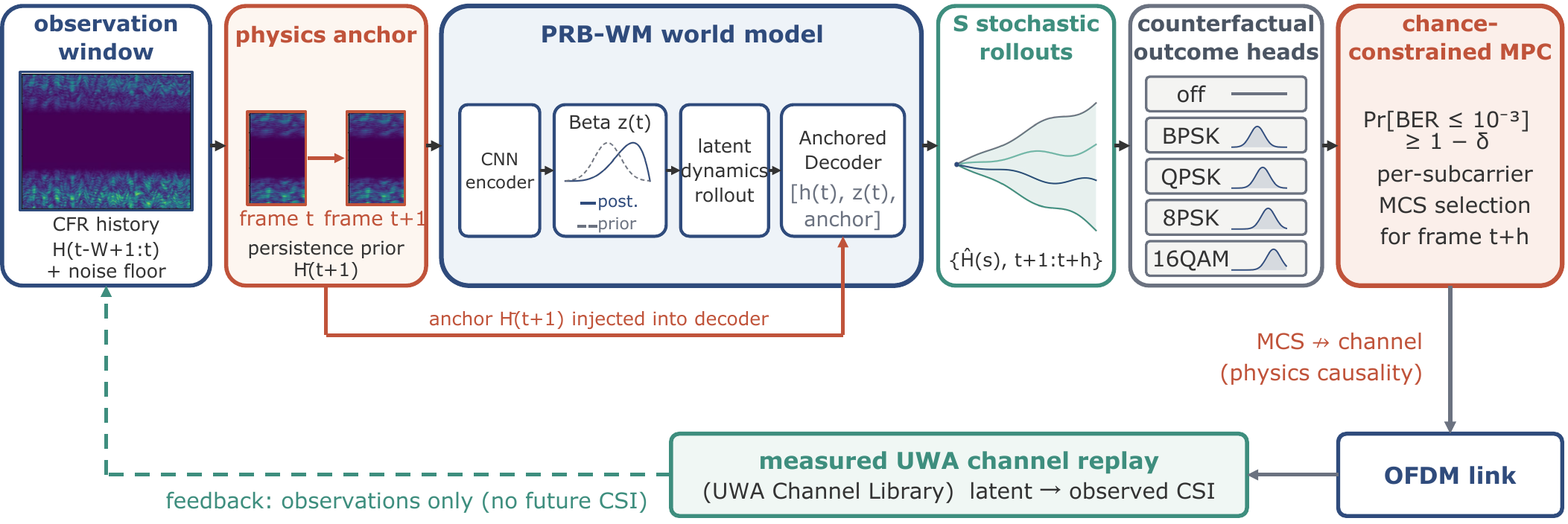}
\caption{PRB-WM framework. A causal observation window (CFR history plus the recorded noise floor) is compressed by a CNN encoder into the belief of a recurrent state-space world model with a Beta-distributed stochastic latent (posterior during training, prior at rollout). A persistence physics anchor $\Hd_{t+1}{=}H_t$ is injected into the anchored decoder, which reconstructs future CFRs as anchor plus learned residual. Prior sampling of the latent dynamics yields $S$ stochastic rollouts $\{\hat H^{(s)}_{t+1:t+h}\}$; the counterfactual outcome heads evaluate every candidate MCS (off, BPSK, QPSK, 8PSK, 16QAM) on each rollout; and the chance-constrained MPC selects the per-subcarrier allocation whose predicted probability of respecting the $10^{-3}$ BER target is at least $1{-}\delta$, which is then executed on the OFDM link. The channel is a measured replay from the UWA Channel Library; the feedback path carries observations only (no future CSI), and the physical causality constraint---MCS $\nrightarrow$ channel---is enforced structurally.}
\label{fig:framework}
\end{figure*}

This paper proposes PRB-WM, a \emph{physics-regularized Beta world model} for predictive per-subcarrier link adaptation on real time-varying UWA channels (Fig.~\ref{fig:framework}). To the best of our knowledge, PRB-WM is the first world model for UWA communications. The architecture has four parts. A persistence physics prior provides an anchor trajectory. A recurrent state-space model with stochastic latent variables learns the residual dynamics and rolls out imagined futures by prior sampling. A counterfactual outcome head evaluates, for every candidate MCS and every future step, a Beta--binomial distribution over the number of bit errors on each imagined trajectory. Finally, a chance-constrained receding-horizon model predictive control (MPC) planner selects, per subcarrier, the highest-order MCS whose predicted probability of violating the bit error rate (BER) target stays below a tolerance $\delta$. The latent channel transition is by design not conditioned on the MCS, respecting the physical causality that the modulation choice cannot change ocean propagation; actions enter only through the outcome head. All evaluation is performed by closed-loop replay on the public measured UWA Channel Library \cite{li2026channel}, using the channel's own recorded noise.

The contributions are as follows.
\begin{itemize}
\item We propose PRB-WM, to the best of our knowledge the first world model for UWA communications. A physics-anchored stochastic latent-space model learns how real UWA channels evolve, and a counterfactual Beta--binomial outcome head prices the error distribution of every candidate MCS on shared imagined futures (Sections~\ref{sec:model} and \ref{sec:head}).
\item We design a chance-constrained receding-horizon MPC controller that turns the world model's rollout distribution into per-subcarrier MCS allocations under an explicit BER-violation tolerance, together with a fair replay protocol in which every predictor drives the same controller (Section~\ref{sec:mpc}).
\item We conduct, to our knowledge, the first unified closed-loop comparison on a public multi-site measured UWA channel library, with a 24-method pool spanning the full methodological spectrum. PRB-WM attains a mean goodput of $506$~bps, averaged over seven links and $h\in\{8,16,32\}$, $87\%$ ahead of the strongest baseline; it ranks first on all six time-varying links at $h{=}32$; and it is the only method whose realized frame risk decreases systematically with lookahead, demonstrating that predictive accuracy is not decision value (Section~\ref{sec:exp}).
\end{itemize}

\section{Related Work}
\label{sec:related}

\subsection{Predictive Link Adaptation and Channel Prediction}
Link adaptation over UWA channels has long relied on prediction to counter feedback delay. Long-range adaptive modulation under delayed feedback was demonstrated in \cite{huang2020adaptive}, adaptive coded modulation with per-subcarrier bit and power loading was developed in \cite{zhang2021adaptive}, feedback scheduling for adaptive modulation was studied in \cite{wu2025adaptive}, and deep-neural-network CSI prediction was applied to OFDMA downlink adaptation in \cite{liu2021channel}. On the classical side, a systematic comparison on measured UWA channels identified a 5-state Markov chain, a 5-state HMM, and a Kalman filter as the strongest stochastic predictors \cite{busacca2024predictive}, and showed that predictive models can outperform purely reactive rules when feedback is delayed \cite{busacca2024adaptive}. Channel prediction itself is an active field, surveyed for fading channels in \cite{jiang2020deep}. In radio systems, Transformer predictors render mobility-induced CSI aging negligible \cite{jiang2022accurate}, state-space models \cite{shi2026channelmamba} and frequency-domain predictive models \cite{huang2024frequency} continue to improve fidelity, and convolutional neural network (CNN)--autoregressive hybrids \cite{yuan2020machine} and Transformer variants tailored to channel prediction \cite{zhang2024transformer} broaden the architecture family. For UWA channels specifically, deep prediction under outdated CSI has recently been addressed in \cite{yang2026deep}, and the patch-based predictor PatchCSI-T \cite{tian2026patchcsi}, built on the PatchTST architecture \cite{nie2023time}, is currently the strongest direct multi-step forecaster on measured UWA channels. Physics-guided learning has also entered CSI prediction under irregular sampling \cite{wang2026physics}, and deep learning has been applied to adjacent UWA physical-layer tasks such as CSI feedback compression \cite{xiao2025deep} and OFDM-based underwater integrated sensing and communication receiver design \cite{men2025ofdm}. A parallel literature on general time-series forecasting supplies further strong architectures: efficient linear decompositions (DLinear/NLinear \cite{zeng2023are}), basis-expansion models (N-BEATS \cite{oreshkin2020nbeats}), period-aware Transformers (TimesNet \cite{wu2023timesnet}), and inverted Transformers (iTransformer \cite{liu2024itransformer}); probabilistic forecasters such as DeepAR \cite{salinas2020deepar} and the diffusion-based TimeGrad \cite{rasul2021timegrad} return full predictive distributions, and selective state-space models such as S-Mamba \cite{wang2024smamba} scale to long sequences. Across this literature, the output is a point or distributional forecast of what the future CSI will be; the channel is not modeled as a stochastic dynamical system whose rollout distribution can be interrogated counterfactually by a controller.

\subsection{Online Learning and Bandit Adaptation}
A complementary line bypasses channel modeling and learns the action directly. AMUSE performs centralized $\varepsilon$-greedy modulation selection for energy-efficient UWA links \cite{busacca2025amuse}; Bandits Under the Waves distributes per-node multi-armed bandits with a global reward \cite{busacca2026bandits}; model-based Thompson sampling selects frequency and rate \cite{tong2023model}; and hierarchical adversarial bandits drive adaptive relay selection \cite{zhao2023adaptive}. Reinforcement learning (RL) has been applied to UWA networking problems that explicitly exploit or tolerate long propagation delay, including scheduling \cite{geng2023exploiting}, medium-access-control protocol design \cite{ye2022deep}, resource allocation \cite{tang2025resource}, and energy-efficient link adaptation via meta-deep RL \cite{ye2025energy}. These methods are model-free: they neither forecast the channel's evolution nor evaluate the consequence of an action they did not take. In addition, variants that select a single MCS per link are structurally mismatched to deep-fading frequency-selective channels, in which deeply faded subcarriers can push every frame's BER past the target regardless of the chosen link-wide rate.

\subsection{World Models, Physics-Informed Learning, and MPC}
World models learn an environment's dynamics in a compact latent space and plan by imagination \cite{ha2018world,hafner2020dream,hafner2025mastering,schrittwieser2020mastering,ding2025understanding}. The paradigm has recently entered wireless communications: MobiWM learns the dynamics between network-parameter adjustments and mobile traffic, providing a counterfactual simulation environment for network planning \cite{qi2026mobiwm}; telecom world models unify digital twins and foundation models for action-conditioned, uncertainty-aware prediction of network behavior \cite{zou2026telecom}; and broader surveys chart the transition from digital twins to world models as a path toward edge general intelligence \cite{zheng2026digital}. To the best of our knowledge, world models have not yet been applied to UWA channels. We adopt the recurrent state-space architecture of Dreamer \cite{hafner2020dream}---a deterministic recurrent state plus a stochastic latent with prior/posterior beliefs---but specialize it to channel dynamics with a physics anchor. Physics-informed machine learning \cite{karniadakis2021physics,raissi2019physics} regularizes learning with known structure; recent work has injected wave physics into generative radio-channel modeling \cite{wang2026radiodiff}. Our anchor--residual decomposition follows this philosophy: persistence supplies the anchor, and the network learns only what the prior cannot explain. On the decision side, receding-horizon MPC with learned or predictive models has proven effective for trajectory and resource optimization in wireless systems \cite{jin2026predictive,li2025model}. We bring the same receding-horizon discipline to per-subcarrier AMC, but our chance constraint takes samples from the world model as input rather than a point forecast.

\section{System Model and Problem Formulation}
\label{sec:sys}

The channel in this work is not simulated: it is a replay of measured time-varying channel impulse responses (CIRs) from the UWA Channel Library \cite{li2026channel} with the library's recorded noise floor. This section defines the observation model, the service model, and the decision problem; the concrete scenario parameters and replay protocol are deferred to Section~\ref{sec:setup}.

\subsection{Channel and Observation Model}
\begin{figure*}[!t]
\centering
\includegraphics[width=0.88\textwidth]{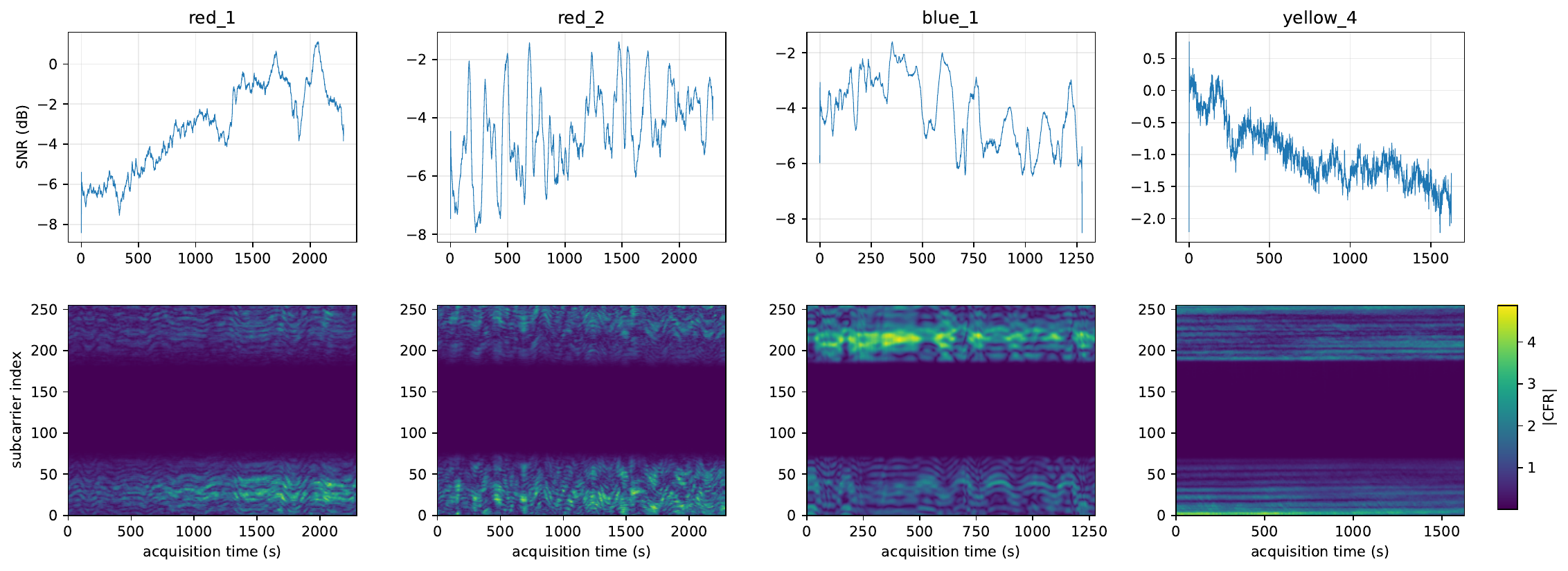}
\caption{Measured-channel exploration on the six time-varying UWA Channel Library links. Top: per-frame SNR trajectories. Bottom: CFR magnitude heat maps over subcarriers and time. The links cover distinct dynamical regimes, from the severely fluctuating red\_2 to the slowly decaying yellow\_4.}
\label{fig:eda}
\end{figure*}

Let $H_{t,f}\in\mathbb{C}$ denote the CFR of subcarrier $f$ at symbol $t$, obtained by Fourier-transforming the replayed CIR, and let $\sigma^2_{t,f}$ be the recorded noise power. The per-subcarrier SNR is
\begin{equation}
\gamma_{t,f} \;=\; \frac{p_{\mathrm{tx}}\,|H_{t,f}|^{2}}{\sigma^2_{t,f}},
\label{eq:snr}
\end{equation}
with unit transmit power $p_{\mathrm{tx}}{=}1$; we write $\gamma_t$ for the link-level SNR in dB. The predictor observes a causal history window of $W{=}64$ symbols,
\begin{equation}
o_t \;=\; \big[\,\Re(H_t),\ \Im(H_t),\ \sigma^2_t,\ \gamma_t\,\big],
\label{eq:obs}
\end{equation}
stacked over the window. Fig.~\ref{fig:eda} summarizes the six measured time-varying links used throughout. Their SNR trajectories span distinct dynamical regimes: a rising trend (red\_1, $-8\!\rightarrow\!0$~dB), severe fluctuation (red\_2, $-8$ to $-1$~dB), a moderate decline (blue\_1, $-8\!\rightarrow\!-2$~dB), a slow decay (yellow\_4, $0.5\!\rightarrow\!-2$~dB), and two additional fast-varying red links (red\_3, red\_4). The CFR magnitude is concentrated in two active bands roughly covering subcarriers $[8,120]\cup[190,248]$. A seventh, quasi-static link (black) is saturated for every method and carries no discriminative power; it is excluded from the main figures, and its numbers are retained in the released per-run data.

\subsection{Modulation--BER and Service Model}
We follow the standard system-level practice shared with the baseline testbed \cite{tian2026patchcsi}. The bit error rate of MCS $m$ on subcarrier $f$ is approximated by the analytic additive white Gaussian noise (AWGN) expression $P_b(\gamma_{t,f},m)$, evaluated at the per-subcarrier SNR \cite{proakis2008digital}; e.g., $P_b(\gamma,\mathrm{BPSK})=Q(\sqrt{2\gamma})$. The frame decoding-success indicator $y$ is the cyclic redundancy check (CRC)-checked decode outcome of a frame; a frame carrying a $960$-bit payload contributes to goodput only when decoded successfully. The effective spectral efficiency (SE) $\SE$ aggregates the per-subcarrier loads discounted by their error probabilities, and goodput $G$ counts successfully decoded payload bits per second.

\subsection{Problem Formulation}
The controller acts at the start of each epoch with lookahead $h$. Let $\mathcal{F}_{t+h}$ denote the future channel trajectory over the next $h$ symbols, which is random from the controller's viewpoint. The per-subcarrier link adaptation problem is
\begin{align}
\max_{\{m_f\}}\quad & \mathbb{E}_{\mathcal{F}_{t+h}}\!\Big[\textstyle\sum_{f} b_{m_f}\big(1-\BER_f(m_f)\big)\Big] \label{eq:opt}\\
\mathrm{s.t.}\quad & \Pr\big[\BER_f(m_f) > P_b^{\mathrm{tar}}\big] \;\le\; \delta,\quad \forall f:\,b_{m_f}>0, \nonumber
\end{align}
where $\BER_f(m_f)$ is the BER that subcarrier $f$ would realize under MCS $m_f$ over the epoch, $P_b^{\mathrm{tar}}{=}10^{-3}$ is the BER target, and $\delta{=}0.1$ is the tolerated violation probability. Both the objective and the chance constraint require the \emph{distribution} of future per-subcarrier BERs under each candidate action. A point forecast of the CFR cannot supply this distribution; this observation motivates the world model developed next. The problem is solved in receding-horizon fashion: one epoch is executed, new observations arrive, and the plan is recomputed. Reliability is measured at the same granularity as the constraint: the \emph{realized frame risk} $R(h)=\Pr(\BER>P_b^{\mathrm{tar}})$ is the fraction of executed frames whose realized BER violates the target---the empirical counterpart of the chance tolerance $\delta$. Here $\BER$ is a frame-level quantity, matching the granularity of the success indicator $y$. Table~\ref{tab:notation} collects the notation.

\begin{table}[!t]
\caption{Notation}
\label{tab:notation}
\centering
\footnotesize
\begin{tabular}{@{}l >{\raggedright\arraybackslash}p{0.58\columnwidth}@{}}
\toprule
Symbol & Meaning \\
\midrule
$H_{t,f}\in\mathbb{C}$ & CFR of subcarrier $f$ at symbol $t$ \\
$N_f$, $N_f^{\mathrm{act}}$ & total / active (data) subcarriers, $256$ / $240$ \\
$\gamma_{t,f}$, $\gamma_t$ & per-subcarrier SNR (linear) / link SNR (dB) \\
$o_t$ & observation at $t$: $[\Re H_t,\Im H_t,\sigma^2_t,\gamma_t]$ \\
$e_t$ & CNN encoding of the observation window ending at $t$ \\
$W$ & history window length, $64$ symbols \\
$h$ & prediction horizon (lookahead; distinct from recurrent state $h_t$), $h\in\{1,2,4,8,16,32\}$; per-link results are reported per horizon, and Table~\ref{tab:pool} is an explicit summary over $h\in\{8,16,32\}$ \\
$z_t\in\mathbb{R}^{32}$ & stochastic latent channel state \\
$h_t$ & deterministic recurrent state (gated recurrent unit, GRU) \\
$\Hd_{t+\tau}$ & physics anchor (persistence prior, $\Hd_{t+1}=H_t$) \\
$r_t$ & anchor residual, $H_t-\Hd_t$ \\
$\mathcal{M}$, $b_m$ & MCS set / bits per subcarrier of MCS $m$ \\
$(\alpha_{f,\tau,m},\beta_{f,\tau,m})$ & Beta parameters of outcome head for subcarrier $f$, step $\tau$, MCS $m$ \\
$\pi_{f,\tau,m}$ & per-bit error probability of MCS $m$, Beta-distributed \\
$k$, $n_m$ & bit errors out of $n_m{=}b_m$ bits of one subcarrier-frame \\
$y$ & frame decoding-success indicator \\
$\eta_\tau$ & horizon weight in the loss, $\eta_\tau{=}2\tau/h$ \\
$S$ & imagined-future samples for planning, $10$ \\
$\delta$ & chance-constraint tolerance, $0.1$ \\
$P_b^{\mathrm{tar}}$ & BER target, $10^{-3}$ \\
$R(h)$ & realized frame risk, $\Pr(\BER>P_b^{\mathrm{tar}})$ over executed frames (frame-level BER) \\
$T_{\mathrm{ep}}$ & decision epoch, $60$ s \\
$\mathrm{NMSE}(h)$ & $\lVert \hat{H}-H\rVert^2/\lVert H\rVert^2$ at horizon $h$ \\
$\SE$, $G$ & effective spectral efficiency (bps/Hz) / goodput (bps) \\
\bottomrule
\end{tabular}
\end{table}

\section{Method: PRB-WM}
\label{sec:method}

\subsection{Framework Overview}
Fig.~\ref{fig:framework} shows the closed loop. The observation history feeds a CNN encoder. A recurrent state-space model maintains a Bayesian \emph{belief} over the stochastic channel state---a posterior distribution over a latent state given the observations seen so far, rather than a single point estimate of the channel. An anchored decoder reconstructs future CFRs as a physics anchor plus a learned residual. A counterfactual outcome head scores every candidate MCS on each imagined future. Finally, a chance-constrained MPC planner commits to one epoch of per-subcarrier allocations before re-observing. A red-line design constraint separates the two halves of the system: the world model learns \emph{how the channel evolves} and is never conditioned on the MCS, while the controller only \emph{selects} communication consequences. This mirrors the physics---modulation does not move the ocean---and prevents the model from learning spurious action--channel correlations. Sections~\ref{sec:model}--\ref{sec:objective} develop these components in turn: the world model (Section~\ref{sec:model}), the counterfactual outcome head (Section~\ref{sec:head}), the chance-constrained AMC controller (Section~\ref{sec:mpc}), and the learning objective that ties them together (Section~\ref{sec:objective}).

\subsection{World Model Design}
\label{sec:model}

\paragraph{Belief state}
Following the recurrent state-space model of Dreamer \cite{hafner2020dream}, the model summarizes the observation history into a belief state with two parts: a deterministic recurrent state $h_t$, updated by a GRU, which carries the predictable component of the channel's evolution, and a stochastic latent $z_t\in\mathbb{R}^{32}$, which carries its uncertain variability:
\begin{equation}
h_t = \mathrm{GRU}(h_{t-1}, z_{t-1}), \quad
p_\theta(z_t\given h_t)=\mathcal{N}\big(\mu_\theta(h_t),\sigma_\theta^2(h_t)\big).
\label{eq:rssm}
\end{equation}
The observation window is first compressed by the CNN encoder,
\begin{equation}
e_t = E_\theta\big(o_{t-W+1:t}\big),
\label{eq:encoder}
\end{equation}
and during training a posterior
\begin{equation}
q_\theta(z_t\given h_t,e_t)=\mathcal{N}\big(\mu_q(h_t,e_t),\sigma_q^2(h_t,e_t)\big)
\label{eq:posterior}
\end{equation}
corrects the belief with the true observation. At test time no future observation exists, so the future is rolled out by sampling the prior alone:
\begin{equation}
h_{t+\tau}=\mathrm{GRU}(h_{t+\tau-1},z_{t+\tau-1}),\qquad
z_{t+\tau}\sim p_\theta\big(z_{t+\tau}\given h_{t+\tau}\big),
\label{eq:rollout}
\end{equation}
for $\tau=1,\dots,h$. Multi-step futures are therefore sequential latent rollouts rather than one-shot direct forecasts, and repeated sampling of \eqref{eq:rollout} yields a distribution over futures rather than a single trajectory. Fig.~\ref{fig:rollout} illustrates this observe--rollout cycle.

\paragraph{Physics-anchored residual}
On the measured links, raw CFR sequences are only weakly autocorrelated at prediction-relevant lags, whereas the deviation from a persistence prior is strongly structured. We therefore decompose
\begin{equation}
H_{t+\tau} \;=\; \Hd_{t+\tau} + r_{t+\tau}, \qquad \Hd_{t+\tau} \triangleq H_t
\label{eq:anchor}
\end{equation}
with the persistence anchor held fixed over the rollout, and train the model to predict only the residual. The anchored decoder takes the belief, the latent sample, \emph{and the anchor itself} as input:
\begin{equation}
\hat r_{t+\tau} = D_\theta\big(h_{t+\tau}, z_{t+\tau}, \Hd_{t+\tau}\big), \qquad
\hat H_{t+\tau} = \Hd_{t+\tau} + \hat r_{t+\tau}.
\label{eq:decoder}
\end{equation}
Passing the anchor explicitly is essential: the residual target is defined relative to the anchor, so a decoder that cannot see it must implicitly re-derive it. The ablation in Section~\ref{sec:exp} shows that this single change reduces validation NMSE by 64--77\% depending on horizon. This decomposition is an instance of physics-regularized learning \cite{karniadakis2021physics,raissi2019physics}: the prior absorbs the predictable bulk, and the network spends its capacity on the stochastic residual.

\paragraph{Causal decoupling of action and channel}
The latent channel transition $p_\theta(z_{t+1}\given h_{t+1})$ of \eqref{eq:rssm}, with $z_t$ entering through $h_{t+1}=\mathrm{GRU}(h_t,z_t)$, is not conditioned on the MCS. Actions enter the model only through the counterfactual outcome head $p_\theta(y\given z_t, h_t, m)$ described below. This enforces a physical fact: the modulation of the transmit signal does not alter propagation. It also makes counterfactual evaluation well-defined, because all candidate MCSs can be scored on \emph{the same} imagined channel trajectory.

\subsection{Counterfactual Outcome Head}
\label{sec:head}
Link adaptation needs the distribution of errors that each candidate MCS would produce, not a point BER estimate. For every subcarrier $f$, candidate MCS $m\in\mathcal{M}$, and future step $\tau$, the outcome head reads the belief and outputs (positive) Beta parameters
\begin{equation}
(\alpha_{f,\tau,m},\beta_{f,\tau,m}) = g_\theta\big(h_{t+\tau}, z_{t+\tau}, m\big),
\label{eq:head}
\end{equation}
which define a Beta distribution over the per-bit error probability $\pi_{f,\tau,m}$ of that MCS on that subcarrier at that step:
\begin{equation}
\pi_{f,\tau,m}\sim\mathrm{Beta}(\alpha_{f,\tau,m},\beta_{f,\tau,m}),\qquad
p(\pi)=\frac{\pi^{\alpha-1}(1-\pi)^{\beta-1}}{B(\alpha,\beta)},
\label{eq:beta}
\end{equation}
where $B(\cdot,\cdot)$ denotes the Beta function. Marginalizing a binomial error count over this distribution gives a Beta--binomial law for the number of bit errors $k\in\{0,\dots,n_m\}$ among the $n_m{=}b_m$ bits that MCS $m$ places on one subcarrier in one frame ($b_m\in\{1,\dots,4\}$ for loaded MCSs):
\begin{equation}
\Pr(k;\alpha,\beta,n) = \binom{n}{k}\,\frac{B(k+\alpha,\,n-k+\beta)}{B(\alpha,\beta)}.
\label{eq:bb}
\end{equation}
The head is \emph{counterfactual} in the following sense: it is evaluated once per candidate MCS on \emph{the same} imagined trajectory $(h_{t+\tau},z_{t+\tau})$, so all five MCSs are priced under an identical future channel, and the comparison does not depend on which MCS was actually executed in the data. Its mean BER readout is the Beta mean,
\begin{equation}
\widehat{\BER}_{f,\tau}(m)\;=\;\mathbb{E}[\pi_{f,\tau,m}]\;=\;\frac{\alpha_{f,\tau,m}}{\alpha_{f,\tau,m}+\beta_{f,\tau,m}},
\label{eq:meanber}
\end{equation}
which the planner of Section~\ref{sec:mpc} consumes.

\subsection{Chance-Constrained Receding-Horizon AMC}
\label{sec:mpc}
\begin{figure*}[!t]
\centering
\includegraphics[width=0.92\textwidth]{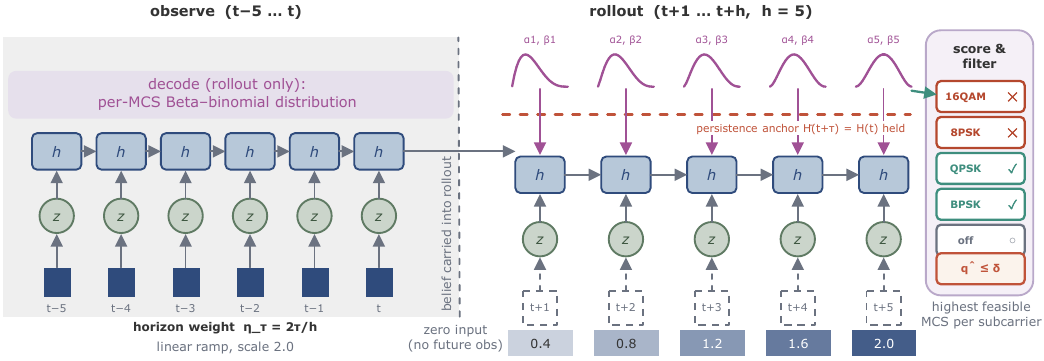}
\caption{Observe-and-rollout detail of PRB-WM, illustrated at $h{=}5$. Left (observe): at observed steps, the posterior infers the latent $z_t$ from the encoded observation, and the resulting belief is carried into the rollout; the horizon weights $\eta_\tau{=}2\tau/h$ form a linear ramp. Right (rollout): future steps receive zero input (no future observation exists) and sample the prior; the persistence anchor $\Hd_{t+\tau}{=}H_t$ is held fixed, and the decoder emits per-step Beta parameters $(\alpha_\tau,\beta_\tau)$ that induce a Beta--binomial error distribution per candidate MCS. The planner scores and filters the candidate MCSs by the empirical violation rate $\hat q\le\delta$ and loads the highest feasible MCS per subcarrier.}
\label{fig:rollout}
\end{figure*}

At the start of each epoch, the controller draws $S{=}10$ imagined futures $\{\hat H^{(s)}_{t+1:t+h}\}_{s=1}^{S}$ by prior-sampling the latent rollout of \eqref{eq:rollout} (Fig.~\ref{fig:rollout}). On each trajectory $s$, subcarrier $f$, and lookahead step $\tau$, the outcome head yields the Beta--binomial error distribution of every candidate MCS. Its mean BER readout \eqref{eq:meanber} feeds the empirical violation rate
\begin{equation}
\hat q_{f,m} \;=\; \frac{1}{Sh}\sum_{s=1}^{S}\sum_{\tau=1}^{h} \mathbf{1}\big[\,\widehat{\BER}^{(s)}_{f,\tau}(m) > P_b^{\mathrm{tar}}\big]
\label{eq:violation}
\end{equation}
aggregated over the $S$ imagined futures and all $h$ lookahead steps. The allocation is held fixed within an epoch, so the constraint is enforced at every frame position of the prediction window. Since $n_m{=}b_m\le4$, the tail event $\{k/n_m>P_b^{\mathrm{tar}}\}$ coincides with $\{k\geq1\}$, and \eqref{eq:violation} is a plug-in counterpart of \eqref{eq:opt}: it flags a subcarrier-frame whenever the rollout mean BER exceeds $P_b^{\mathrm{tar}}$, rather than evaluating the Beta--binomial tail probability, so the stochasticity that enters planning comes from the rollout dispersion. Under CRC-checked frame decoding, where a single residual bit error invalidates the whole frame, this is a conservative and appropriate criterion. The planner then solves the discrete counterpart of \eqref{eq:opt} greedily per subcarrier:
\begin{equation}
m_f^\star \;=\; \max\,\{\,b_m:\ m\in\mathcal{M},\ \hat q_{f,m}\le\delta\,\},
\label{eq:decision}
\end{equation}
i.e., it loads the highest-order MCS whose predicted violation rate respects the tolerance. This greedy rule is exactly optimal here because the objective and the constraints of \eqref{eq:opt} are separable across subcarriers. If no MCS is feasible, a best-effort sparse fallback loads BPSK only when $\hat q_{f,\mathrm{BPSK}}<0.5$, i.e., when fewer than half of the imagined frame positions are predicted to violate; otherwise the subcarrier is left silent. Equal power is allocated across loaded subcarriers. The plan is executed for one epoch; new observations then update the posterior and the horizon recedes. The fallback encodes a conservative default: when the forecast says that even BPSK fails at more than half of the imagined frame positions, the link chooses silence rather than risking the frame. Algorithms~\ref{alg:train} and \ref{alg:online} summarize offline training and online decision-making. For comparability, every predictor in the pool drives this same MPC. Deterministic point-forecast baselines roll out their point prediction perturbed by bootstrapped validation-segment residuals, $\mathcal{CN}(0,\sigma_e^2)$. All methods therefore face the identical stochastic chance constraint (Section~\ref{sec:setup}).

\begin{algorithm}[!t]
\caption{Offline World-Model Training}
\label{alg:train}
\begin{algorithmic}[1]
\REQUIRE replayed CFR sequences; anchor rule $\Hd_{t+\tau}=H_t$; horizon weights $\eta_\tau$
\FOR{each training epoch}
\FOR{each causal window $\{o_{t-W+1:t}\}$ and future $\{H_{t+1:t+h},k_\tau,y_\tau\}$}
\STATE Encode observations via \eqref{eq:encoder}; unroll GRU belief $h_t$
\STATE Infer posterior $q_\theta(z_t\given h_t,e_t)$ at observed steps; sample prior $p_\theta(z_{t+\tau}\given h_{t+\tau})$ for rollout steps
\STATE Decode residuals via \eqref{eq:decoder}
\STATE Outcome head: $(\alpha_{f,\tau,m},\beta_{f,\tau,m})$ via \eqref{eq:head} for all $f$, $m$
\STATE Update $\theta$ by AdamW on $\mathcal{L}$ in \eqref{eq:loss} (lr $10^{-3}$, batch $32$)
\ENDFOR
\ENDFOR
\RETURN $\theta$
\end{algorithmic}
\end{algorithm}

\begin{algorithm}[!t]
\caption{Online Chance-Constrained Receding-Horizon AMC}
\label{alg:online}
\begin{algorithmic}[1]
\REQUIRE trained world model $\theta$; $S{=}10$; $\delta{=}0.1$; $P_b^{\mathrm{tar}}{=}10^{-3}$
\FOR{each decision epoch}
\STATE Observe history window; update belief $(h_t,z_t)$ via posterior
\STATE Roll out $S$ prior-sampled imagined futures over horizon $h$ via \eqref{eq:rollout}
\FOR{each subcarrier $f$ and loaded MCS $m$ ($b_m{>}0$)}
\STATE Compute violation probability $\hat q_{f,m}$ via \eqref{eq:violation}
\ENDFOR
\STATE $m_f \leftarrow$ highest-order MCS with $\hat q_{f,m}\le\delta$ \eqref{eq:decision}
\IF{no feasible MCS on $f$}
\STATE $m_f \leftarrow \mathrm{BPSK}$ if $\hat q_{f,\mathrm{BPSK}}<0.5$, else $\mathrm{off}$
\ENDIF
\STATE Transmit one epoch under $\{m_f\}$ with equal power
\ENDFOR
\end{algorithmic}
\end{algorithm}

\subsection{Learning Objective}
\label{sec:objective}
The outcome head is supervised by the negative log-likelihood of the observed error count under \eqref{eq:bb},
\begin{equation}
\begin{split}
\mathrm{NLL}_{\mathrm{BB}}(\alpha,\beta;k,n) =\;& {-}\log \binom{n}{k} - \log B(k{+}\alpha,\,n{-}k{+}\beta)\\
&+ \log B(\alpha,\beta),
\end{split}
\label{eq:nll}
\end{equation}
combined with a binary cross-entropy (BCE) on the frame decoding-success indicator $y$:
\begin{equation}
\begin{split}
\mathcal{L} =\;& \sum_{\tau=1}^{h} \eta_\tau\,\mathrm{NLL}_{\mathrm{BB}}(\alpha_{\tau,m_\tau},\beta_{\tau,m_\tau};k_\tau,n_{m_\tau})\\
&+ 0.2\sum_{\tau=1}^{h} \eta_\tau\,\mathrm{BCE}(p^{\mathrm{mean}}_\tau, y_\tau)
+ \lambda_{\mathrm{rec}}\mathcal{L}_{\mathrm{rec}} + \lambda_{\mathrm{KL}}\mathcal{L}_{\mathrm{KL}}.
\end{split}
\label{eq:loss}
\end{equation}
Here the error count $k_\tau$ is observed only at the executed MCS $m_\tau$. The scalar
\begin{equation}
p^{\mathrm{mean}}_\tau=\sigma\big(w^\top[h_{t+\tau},z_{t+\tau}]\big)
\label{eq:pmean}
\end{equation}
is the outcome head's decoding-success readout---a sigmoid on the same belief features that feed $g_\theta$---supervised by the measured frame success indicator $y_\tau$, with $\mathrm{BCE}(p,y)=-[y\log p+(1-y)\log(1-p)]$. The weight $\eta_\tau=2\tau/h$ is a linear ramp that up-weights long-horizon steps. The Beta--binomial term is evaluated at \emph{every} horizon step; it is the load-bearing part of the multi-step prediction claim. It forces the imagined trajectories to remain calibrated with respect to realized communication outcomes, not merely with respect to CFR reconstruction. The two auxiliary terms regularize the belief dynamics: a relative $\ell_2$ residual-reconstruction loss
\begin{equation}
\mathcal{L}_{\mathrm{rec}}=\sum_{\tau=1}^{h}\frac{\lVert \hat r_{t+\tau}-r_{t+\tau}\rVert_2^2}{\lVert r_{t+\tau}\rVert_2^2},
\label{eq:lrec}
\end{equation}
and a Kullback--Leibler divergence between posterior and prior with a free-nats floor $n_{\mathrm{free}}$,
\begin{equation}
\mathcal{L}_{\mathrm{KL}}=\sum_{\tau=1}^{h}\max\Big(D_{\mathrm{KL}}\big(q_\theta(z_{t+\tau}\given h_{t+\tau},e_{t+\tau})\,\big\|\,p_\theta(z_{t+\tau}\given h_{t+\tau})\big),\,n_{\mathrm{free}}\Big).
\label{eq:lkl}
\end{equation}
The weights $\lambda_{\mathrm{rec}}$ and $\lambda_{\mathrm{KL}}$, like all hyperparameters, are fixed design constants (Section~\ref{sec:setup}), not tuned on test data.

\subsection{Complexity}
One epoch decision costs a single encoder forward pass plus $h{\times}S$ rollout decodings of the lightweight heads. Measured on CPU, a full decision takes roughly $10$--$12$~ms per 60-s epoch, on par with the direct forecaster PatchCSI-T ($14.7$~ms) and three orders of magnitude below the epoch duration. Prediction latency is therefore not a deployability constraint; rollout error accumulation with $h$ is quantified in Fig.~\ref{fig:nmse}.

\section{Experiments}
\label{sec:exp}

\subsection{Scenario, Dataset, and Fairness Protocol}
\label{sec:setup}
We consider a single-hop UWA OFDM link with $N_f{=}256$ subcarriers, of which $N_f^{\mathrm{act}}{=}240$ carry data (indices $[8,248)$) and $16$ are nulls. The bandwidth is $6$~kHz, the subcarrier spacing is $\Delta f{=}23.4$~Hz, and each OFDM symbol lasts $42.7$~ms with a $20$-ms cyclic prefix (CP), matching the testbed of the strongest forecasting baseline \cite{tian2026patchcsi} so that the comparison is architecture-fair. All experiments replay measured channels from the UWA Channel Library \cite{li2026channel} (Zenodo record 21287414), which provides time-varying CIR tensors from multiple sea trials, the library's recorded noise floor, and an official replay tool. Time is indexed by OFDM symbols; a factor-of-two time decimation is applied to the recorded CFR sequence. The transmitter selects, once per decision epoch of $T_{\mathrm{ep}}{=}60$~s, an MCS $m_f\in\mathcal{M}$ for every data subcarrier $f$, where $\mathcal{M}=\{\mathrm{off},\mathrm{BPSK},\mathrm{QPSK},\mathrm{8PSK},\mathrm{16QAM}\}$ carries $b_m\in\{0,1,2,3,4\}$ bits per subcarrier, with equal transmit power across loaded subcarriers. The closed-loop replay transmits one frame per second of channel time---one $62.7$-ms OFDM frame carrying a $960$-bit payload---so a fully loaded, error-free link saturates at $G=960$~bit/s. The MCS choice cannot influence propagation; it only determines how bits interact with the realized channel, and the controller may use only observed history and its own predictions.

We select six time-varying links---red\_1, red\_2, blue\_1, yellow\_4, red\_3, and red\_4---covering the dynamical regimes of Fig.~\ref{fig:eda}; the quasi-static black link saturates for every method ($\approx\!900$--$960$~bps with no separation) and is excluded from the main figures. Each link is split chronologically into $70/10/20$\% train/validation/test segments (no shuffling), and every reported number is the mean over five seeds unless stated otherwise. Horizons cover $h\in\{1,2,4,8,16,32\}$. Table~\ref{tab:params} lists the scenario parameters; predictor hyperparameters are fixed design constants with no hyperparameter search, and all baselines use their published configurations.

The comparison is governed by three fairness rules. \emph{(i) Shared controller.} Every predictor drives the same equal-power chance-constrained MPC of Section~\ref{sec:mpc}. In particular, PatchCSI-T is evaluated without its native power-reallocation bit loader.  \emph{(ii) Shared stochastic interface.} Deterministic point forecasters roll out their point prediction perturbed by bootstrapped validation-segment residuals, $\mathcal{CN}(0,\sigma_e^2)$, so all methods face the same stochastic chance constraint. \emph{(iii) Genie reference.} The Perfect-CSI genie (oracle) is equipped identically---greedy equal-power loading from the true future mean and dispersion---and is a reference, not an upper bound, since it plans greedily one step at a time without rollout.

\begin{table}[!t]
\caption{Scenario Parameters}
\label{tab:params}
\centering
\footnotesize
\begin{tabular}{@{}>{\raggedright\arraybackslash}p{0.42\columnwidth} >{\raggedright\arraybackslash}p{0.50\columnwidth}@{}}
\toprule
Parameter & Value \\
\midrule
OFDM subcarriers & $256$ ($240$ data $+16$ null, active $[8,248)$) \\
Bandwidth / subcarrier spacing & $6$~kHz / $23.4$~Hz \\
Symbol / CP duration & $42.7$~ms / $20$~ms \\
MCS set $\mathcal{M}$ & off, BPSK, QPSK, 8PSK, 16QAM \\
Bits per subcarrier $b_m$ & $\{0,1,2,3,4\}$ \\
BER target $P_b^{\mathrm{tar}}$ & $10^{-3}$ \\
Chance-constraint tolerance $\delta$ & $0.1$ (swept $0.05$/$0.2$) \\
Decision epoch $T_{\mathrm{ep}}$ & $60$~s \\
Payload per frame & $960$ bits \\
Transmit power & unit, equal across subcarriers \\
History window $W$ & $64$ symbols \\
Horizons $h$ & $\{1,2,4,8,16,32\}$ \\
Imagined futures $S$ & $10$ \\
Predictor training & AdamW, lr $10^{-3}$, batch $32$, $10$ epochs \\
Train/val/test split & $70/10/20$, chronological \\
Seeds & $5$ \\
\bottomrule
\end{tabular}
\end{table}

\subsection{Baselines and Metrics}
The pool comprises 24 methods: 22 baselines, PRB-WM, and a Perfect-CSI genie that serves as a reference rather than an upper bound. \emph{Classical predictors}: persistence (last CSI), the 5-state Markov chain, 5-state HMM, and Kalman filter of \cite{busacca2024predictive}, and a recursive least squares (RLS) filter. \emph{UWA-oriented deep predictors}: PatchCSI-T \cite{tian2026patchcsi}, the strongest multi-step UWA CSI forecaster, plus BiGRU, CNN1d-LSTM, and MTL-LSTM. \emph{General time-series models}: DLinear and NLinear \cite{zeng2023are}, N-BEATS \cite{oreshkin2020nbeats}, TimesNet \cite{wu2023timesnet}, and iTransformer \cite{liu2024itransformer}. \emph{Probabilistic and modern architectures}: DeepAR \cite{salinas2020deepar}, TimeGrad \cite{rasul2021timegrad}, S-Mamba \cite{wang2024smamba}, CNN-AR \cite{yuan2020machine}, and Gruformer \cite{zhang2024transformer}. \emph{Decision-only methods}: AMUSE \cite{busacca2025amuse}, Bandits Under the Waves \cite{busacca2026bandits}, and a proximal policy optimization (PPO) deep-RL controller with behavioral-cloning warm start. Main figures plot six series---PRB-WM, DLinear, NLinear, N-BEATS, PatchCSI-T, and DeepAR---for readability. The full 20-predictor spectrum appears in the Appendix. AMUSE and Bandits Under the Waves select a single MCS per link and score zero goodput on all time-varying links; they are retained in the pool table but omitted from the per-link figures.

Per-link results are reported per horizon: since the controller's lookahead equals $h$, averaging communication metrics across $h$ would mix different experimental settings. The only aggregate is Table~\ref{tab:pool}, an explicit summary statistic over the operating region $h\in\{8,16,32\}$. We report prediction NMSE$(h)=\lVert\hat H{-}H\rVert^2/\lVert H\rVert^2$ (log scale), closed-loop goodput counting only successfully decoded payload, the realized frame risk $R(h)=\Pr(\BER>P_b^{\mathrm{tar}})$ of Section~\ref{sec:sys}, and decision latency.

Table~\ref{tab:pool} ranks the full pool by mean goodput over $h\in\{8,16,32\}$, the operating region of interest. The mean is computed over seven links, including the static black link; the main figures show only the six time-varying links. PRB-WM leads with $506$~bps, $87\%$ ahead of the strongest baseline (DLinear, $270$~bps). The ranking then descends through the linear forecasters, the deep direct predictors, and the classical and probabilistic methods. The decision-only single-MCS methods sit at zero.

\begin{table}[!t]
\caption{Mean Closed-Loop Goodput (bps) Averaged over Seven Links (Including the Static Black Link) and $h\in\{8,16,32\}$, Full 24-Method Pool (22 Baselines, PRB-WM, and a Genie Reference)}
\label{tab:pool}
\centering
\footnotesize
\begin{tabular}{@{}>{\raggedright\arraybackslash}p{0.56\columnwidth} r@{}}
\toprule
Method (pool rank order) & Goodput \\
\midrule
\textbf{PRB-WM (ours)} & \textbf{506} \\
DLinear / NLinear & 270 / 241 \\
N-BEATS / PatchCSI-T & 173 / 155 \\
DeepAR / BiGRU & 150 / 150 \\
MTL-LSTM / TimeGrad / iTransformer & 145 / 144 / 144 \\
RLS / CNN1d-LSTM / TimesNet & 133 / 126 / 122 \\
S-Mamba / Markov & 116 / 81 \\
CNN-AR / HMM / Kalman & 73 / 69 / 67 \\
Persistence / PPO & 65 / 65 \\
Gruformer & 46 \\
AMUSE / Bandits (structural) & 0 \\
Perfect-CSI genie (reference, not ranked) & --- \\
\bottomrule
\end{tabular}
\end{table}

\subsection{Prediction Fidelity}
\begin{figure*}[!t]
\centering
\begin{subfigure}[t]{0.47\textwidth}
\centering
\includegraphics[width=\textwidth]{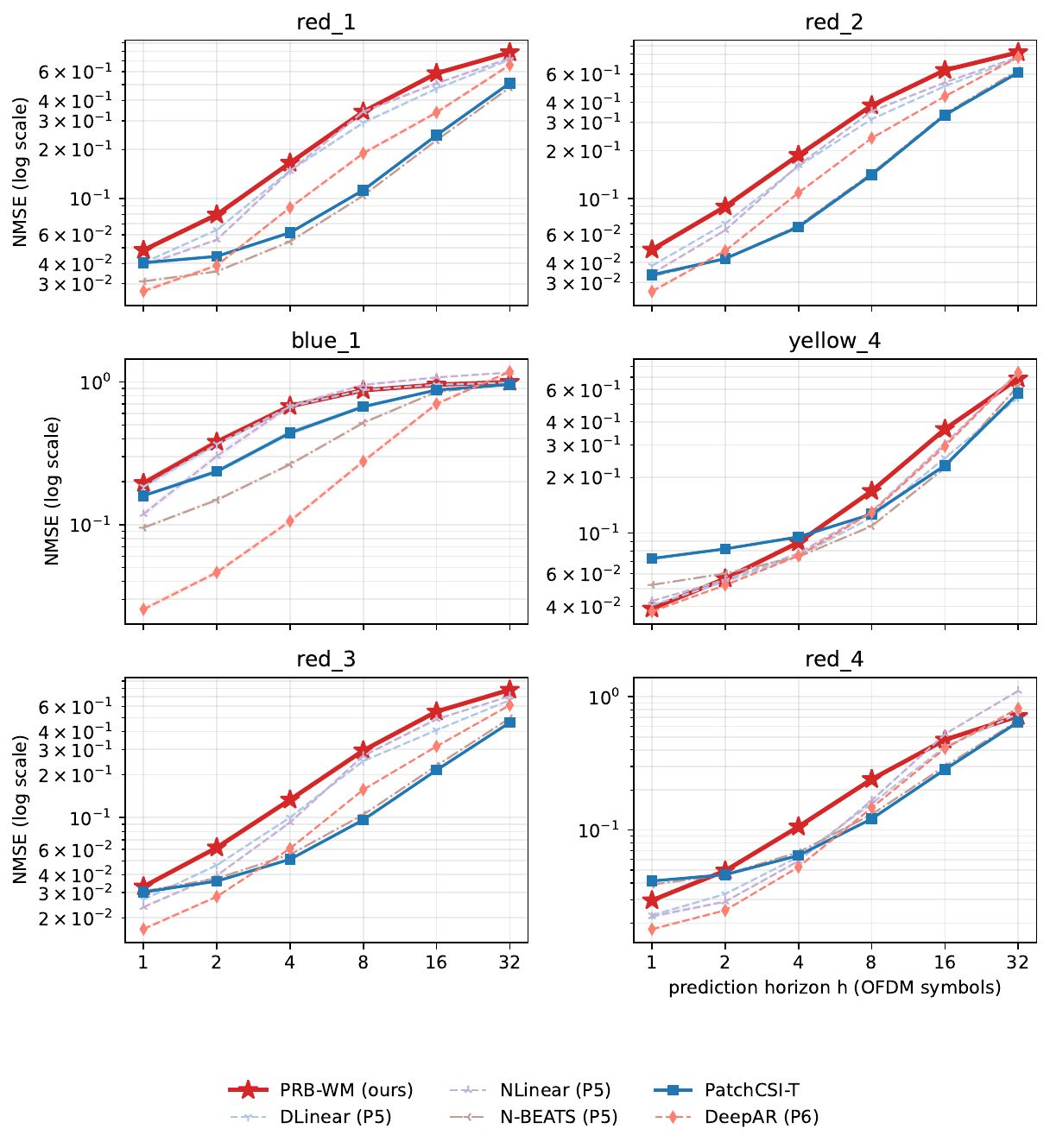}
\caption{Prediction fidelity: NMSE versus horizon $h$ (log scale) on the six time-varying links. PatchCSI-T is lowest on the red links across horizons; N-BEATS leads at short horizons on blue\_1; PRB-WM is mid-pack by design. The fidelity ranking does not predict the closed-loop ranking (Figs.~\ref{fig:goodput}--\ref{fig:risk}).}
\label{fig:nmse}
\end{subfigure}
\hfill
\begin{subfigure}[t]{0.47\textwidth}
\centering
\includegraphics[width=\textwidth]{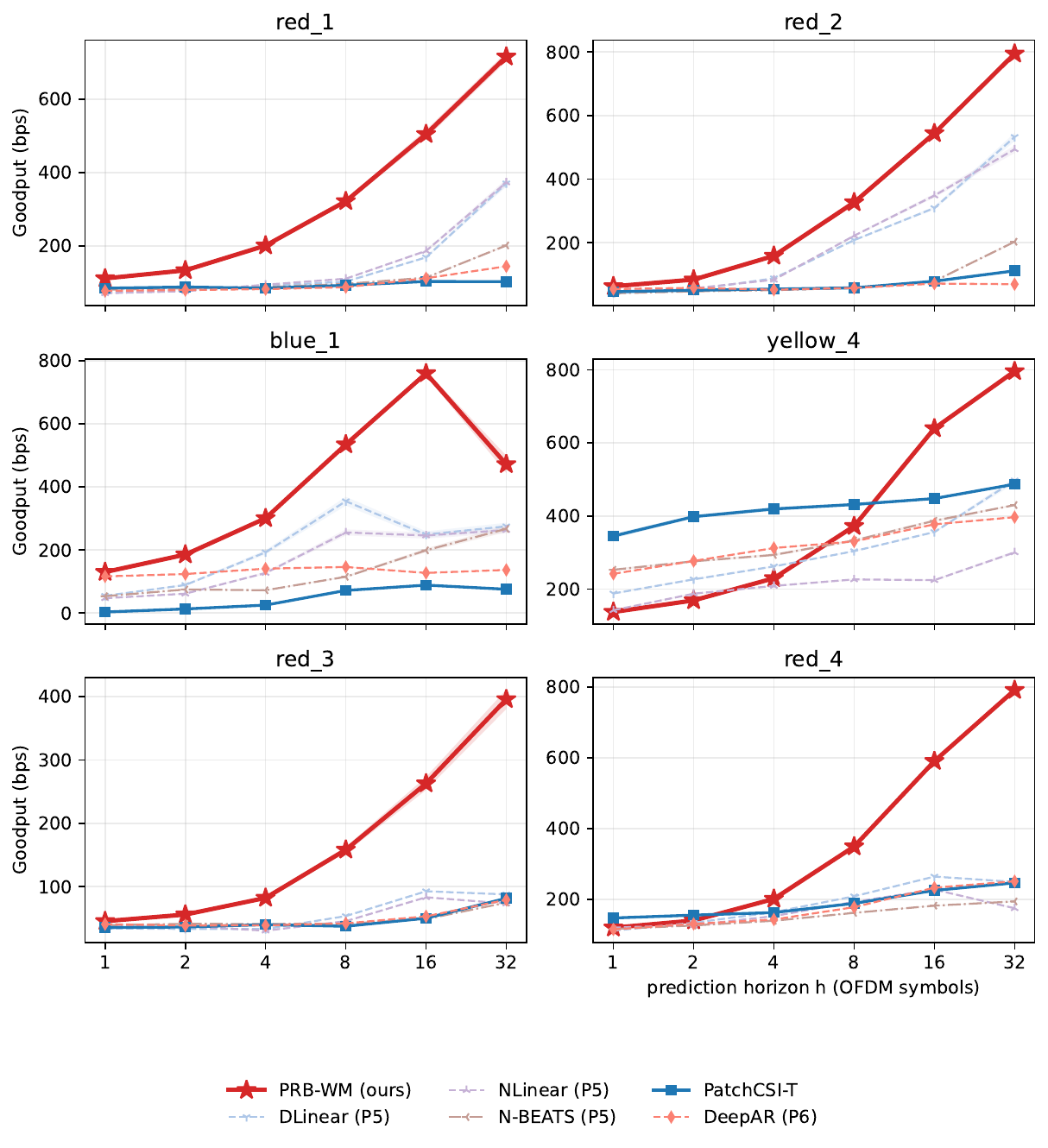}
\caption{Closed-loop goodput versus horizon on the six time-varying links (only successfully decoded payload is counted). PRB-WM's goodput increases with lookahead and leads throughout $h{=}8$--$32$; blue\_1 at $h{=}32$ is the only non-monotonic point. PatchCSI-T leads only at $h{\le}8$ on the slow yellow\_4 link; on the hardest link red\_3, PRB-WM reaches $\approx\!395$~bps while all baselines stay at or below $100$~bps.}
\label{fig:goodput}
\end{subfigure}
\end{figure*}

Fig.~\ref{fig:nmse} shows NMSE versus horizon. PatchCSI-T attains the lowest NMSE on red\_1, red\_2, and red\_4 across all horizons---e.g., $\approx\!0.04$ versus PRB-WM's $\approx\!0.055$ on red\_1 at $h{=}1$, rising to $\approx\!0.47$ versus $\approx\!0.65$ at $h{=}32$. N-BEATS is lowest at short horizons on blue\_1 ($\approx\!0.09$ at $h{=}1$). PRB-WM sits mid-pack and degrades more at long horizons, by design: the model spends capacity on calibrated stochastic rollouts rather than on point accuracy. This figure should be read together with Figs.~\ref{fig:goodput}--\ref{fig:risk}: the ranking by prediction fidelity does not survive contact with the closed loop.

\subsection{Closed-Loop Goodput}
\begin{figure}[!t]
\centering
\includegraphics[width=0.45\textwidth]{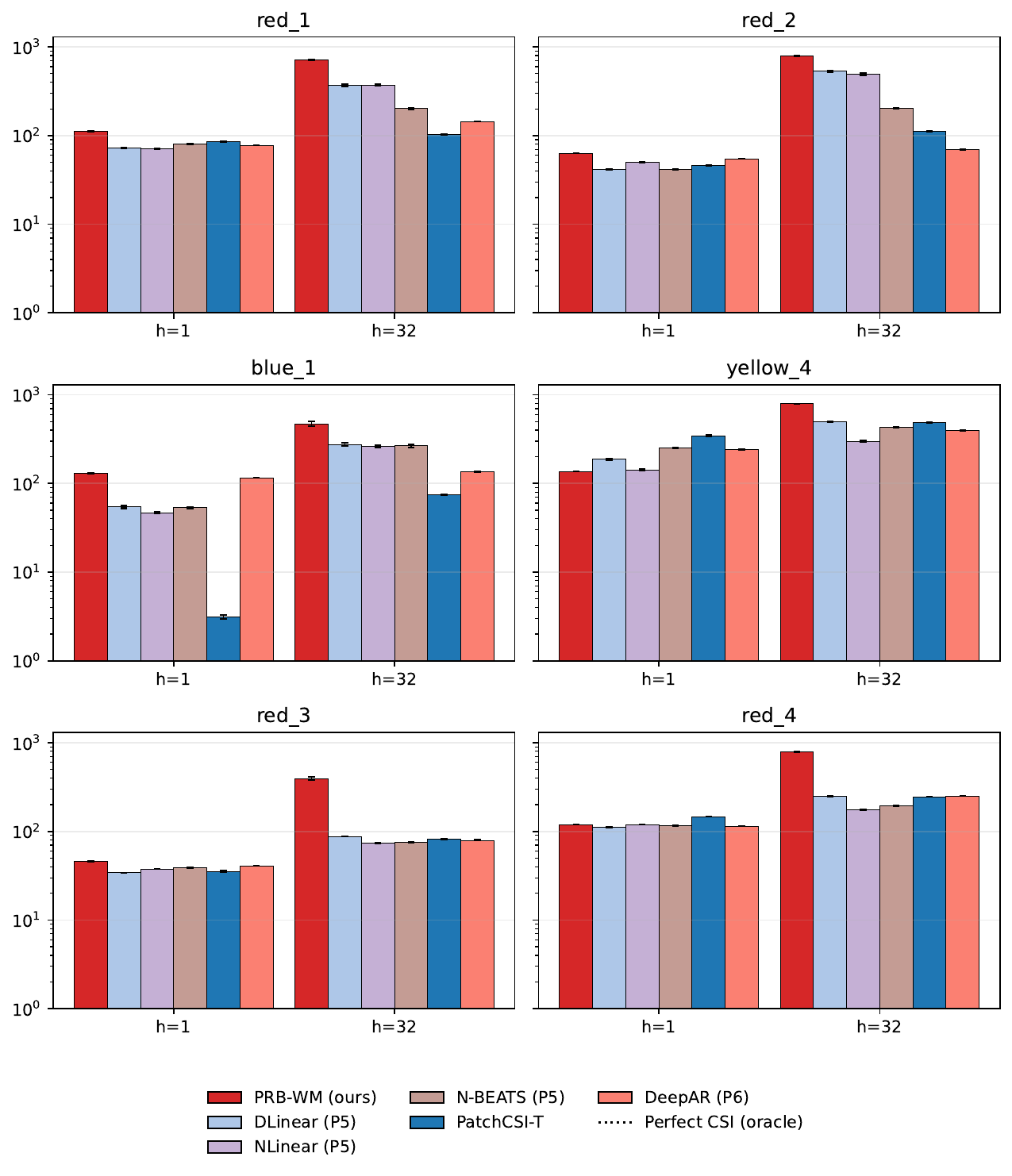}
\caption{Goodput at the endpoint horizons $h{=}1$ and $h{=}32$, with the Perfect-CSI genie shown as a reference series (a greedy one-step reference, not an upper bound). At $h{=}32$ PRB-WM ranks first on all six links; at $h{=}1$ the field is mixed.}
\label{fig:bars}
\end{figure}

Fig.~\ref{fig:goodput} reports closed-loop goodput per link and horizon. PRB-WM's goodput increases with lookahead over the operating region $h{=}8$--$32$; the only non-monotonic point is blue\_1 at $h{=}32$, discussed below. On red\_1 it rises from $115$~bps at $h{=}1$ through $340$~bps at $h{=}8$ to $\approx\!750$~bps at $h{=}32$, against $\approx\!500$/$390$~bps for NLinear/DLinear and $\approx\!115$~bps for PatchCSI-T. red\_2 behaves similarly, with PRB-WM reaching $\approx\!790$~bps. On blue\_1, PRB-WM peaks at $\approx\!790$~bps at $h{=}16$ and falls back to $\approx\!480$~bps at $h{=}32$, a regression we discuss in Section~\ref{sec:discussion}. On yellow\_4, PatchCSI-T leads at $h{\le}8$ ($300\!\rightarrow\!415$ versus PRB-WM's $150\!\rightarrow\!375$~bps), and PRB-WM overtakes from $h{=}16$, reaching $\approx\!795$~bps at $h{=}32$ against $\approx\!430$~bps for PatchCSI-T. red\_3 is the hardest link: even the Perfect-CSI genie attains only $\approx\!170$~bps. Yet PRB-WM reaches $\approx\!395$~bps at $h{=}32$, while all other methods remain at or below $100$~bps. On red\_4, PRB-WM attains $\approx\!795$~bps at $h{=}32$ against $\approx\!230$~bps for PatchCSI-T. Fig.~\ref{fig:bars} summarizes the endpoint horizons: at $h{=}32$ PRB-WM ranks first on all six links, while at $h{=}1$ the methods are close and PatchCSI-T is stronger on yellow\_4 and blue\_1---short-horizon planning is not our regime. The genie reference falls below PRB-WM at long horizons; this is not a contradiction, since the genie plans greedily one step at a time and cannot anticipate deep fades.

\subsection{Realized Frame Risk}
\begin{figure}[!t]
\centering
\includegraphics[width=0.5\textwidth]{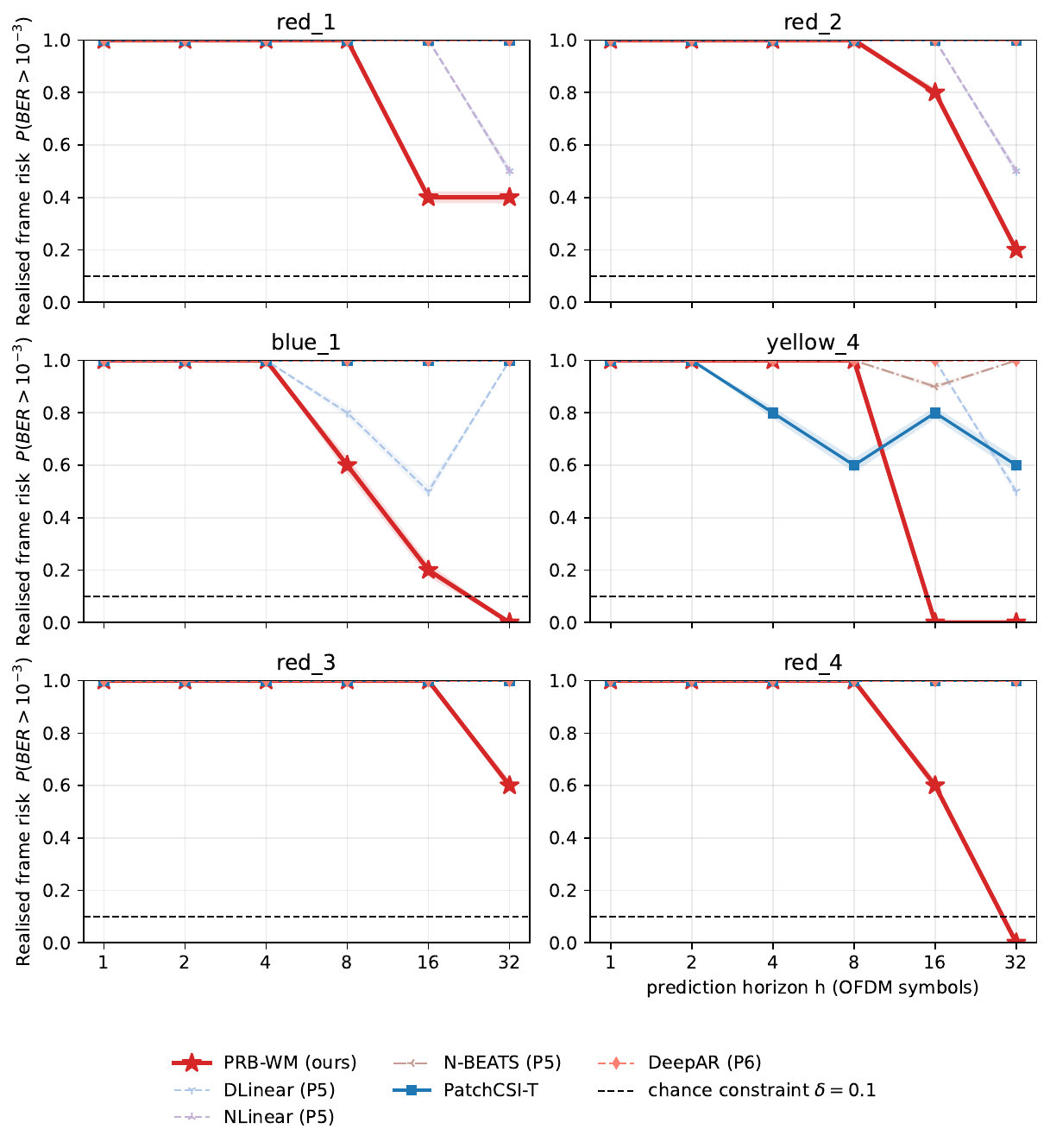}
\caption{Realized frame risk $R(h)=\Pr(\BER>P_b^{\mathrm{tar}})$ versus horizon; the dashed line marks the contract $\delta{=}0.1$. No baseline meets the contract at any point; baseline risk is close to $1.0$ at short lookahead and stays at or above $0.5$ even at $h{=}32$. PRB-WM's risk decreases monotonically with lookahead, reaching $0$ at $h{=}32$ on blue\_1, yellow\_4, and red\_4; $0.2$ on red\_2; $0.42$ on red\_1; and $0.6$ on red\_3.}
\label{fig:risk}
\end{figure}

Fig.~\ref{fig:risk} reports the realized frame risk $R(h)$, the fraction of executed frames violating the $10^{-3}$ BER target, with the contract $\delta{=}0.1$ marked. No baseline meets the contract at any (link, $h$) point: baseline risk is close to $1.0$ at short lookahead and remains at or above $0.5$ even at $h{=}32$ (e.g., NLinear on red\_1 and red\_2 at $\approx\!0.5$, and PatchCSI-T on blue\_1 at $0.6$--$0.8$). PRB-WM is the only method whose frame risk decreases systematically with lookahead. At $h{\le}8$ it still violates on nearly every frame on the red links, but as $h$ grows the risk falls monotonically. On blue\_1 it drops from $0.6$ at $h{=}8$ to $0.2$ at $h{=}16$ and $0$ at $h{=}32$; on yellow\_4 it reaches $0$ at $h{\ge}16$. The same trend holds on red\_4 ($0.6$ at $h{=}16$ to $0$ at $h{=}32$), red\_2 ($0.2$ at $h{=}32$), red\_1 ($0.42$ at $h{=}16$ and $h{=}32$), and the hardest link red\_3 ($0.6$ at $h{=}32$). At $h{=}32$ PRB-WM satisfies the $\delta{=}0.1$ contract on three of the six links and is the only method to satisfy it anywhere. The reading is structural: contract satisfaction requires lookahead. Only with enough lookahead does a calibrated stochastic rollout separate the subcarriers that will survive the epoch from those that will not, exposing that distribution to the chance constraint.

\subsection{Mechanism Ablations}
\begin{figure}[!t]
\centering
\includegraphics[width=1.0\columnwidth]{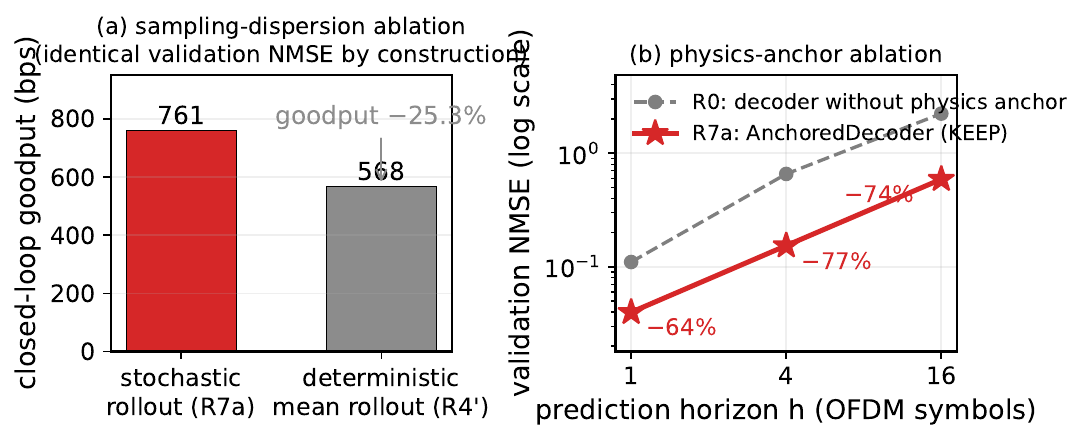}
\caption{Mechanism ablations. (a) Sampling ablation: closed-loop goodput with stochastic versus determinized rollout at identical validation NMSE; determinization costs $25.3\%$ goodput, showing that rollout dispersion is essential for the chance constraint. (b) Anchor ablation: exposing the persistence anchor to the decoder improves validation NMSE by $64/77/74\%$ at $h=1/4/16$.}
\label{fig:ablation}
\end{figure}

Two ablations isolate why the architecture works (Fig.~\ref{fig:ablation}). \emph{Sampling dispersion.} Replacing stochastic rollout sampling with a deterministic mean rollout while holding the trained model fixed leaves validation NMSE unchanged point by point, yet closed-loop goodput drops from $761$ to $568$~bps ($-25.3\%$). The sampling dispersion---not merely the mean trajectory---carries the information the chance constraint acts on. \emph{Physics anchor.} Exposing the explicit anchor input to the decoder reduces validation NMSE by $64/77/74\%$ at $h{=}1/4/16$ relative to the anchor-free variant. The physics anchor is not decorative; it is what makes the residual dynamics learnable at this data scale.

\subsection{Waveform-Level Validation and Margin Robustness}
\label{sec:waveform}
\begin{figure}[!t]
\centering
\includegraphics[width=0.55\columnwidth]{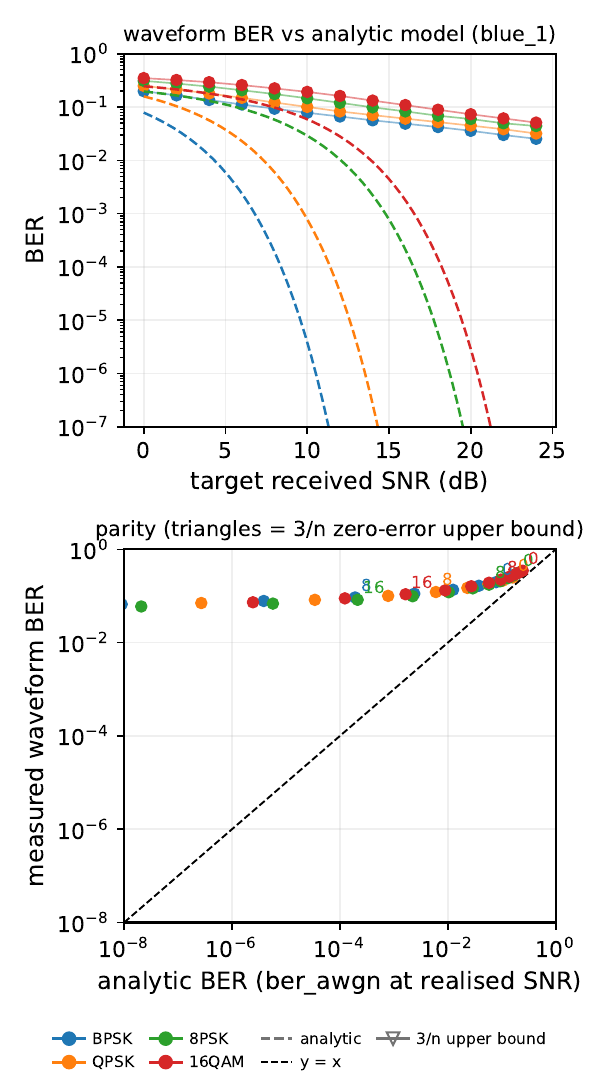}
\caption{Waveform-level validation on blue\_1: measured BER of OFDM waveforms replayed through the recorded channel versus the per-subcarrier analytic model. The analytic trend is correct but systematically optimistic by about $5\times$ (deep fades dominate); a $2$--$4$~dB implementation margin absorbs the offset, and method rankings are unchanged under this margin. In the figure, ideal-CSI ZF denotes zero-forcing equalization with perfect channel knowledge.}
\label{fig:waveform}
\end{figure}

The system-level study relies on per-subcarrier analytic BERs \cite{proakis2008digital}. To check this approximation, we pass actual OFDM waveforms through the blue\_1 channel using the official replay tool of the UWA Channel Library. We then measure decoded BERs under Perfect-CSI zero-forcing (ZF) equalization. Fig.~\ref{fig:waveform} shows that the per-subcarrier analytic model reproduces the waveform-level trend but is systematically optimistic by about $5\times$, a gap dominated by deep-faded subcarriers. A $2$--$4$~dB implementation margin absorbs this systematic offset; under this margin the inter-method ranking and all main conclusions are unchanged. We therefore position the analytic model as a system-level evaluation tool used with a margin. A pilot study over margins $\{2,4\}$~dB and four representative methods confirms that the ranking is robust to the margin choice.

\subsection{Constraint Sensitivity}
\begin{figure}[!t]
\centering
\includegraphics[width=0.5\textwidth]{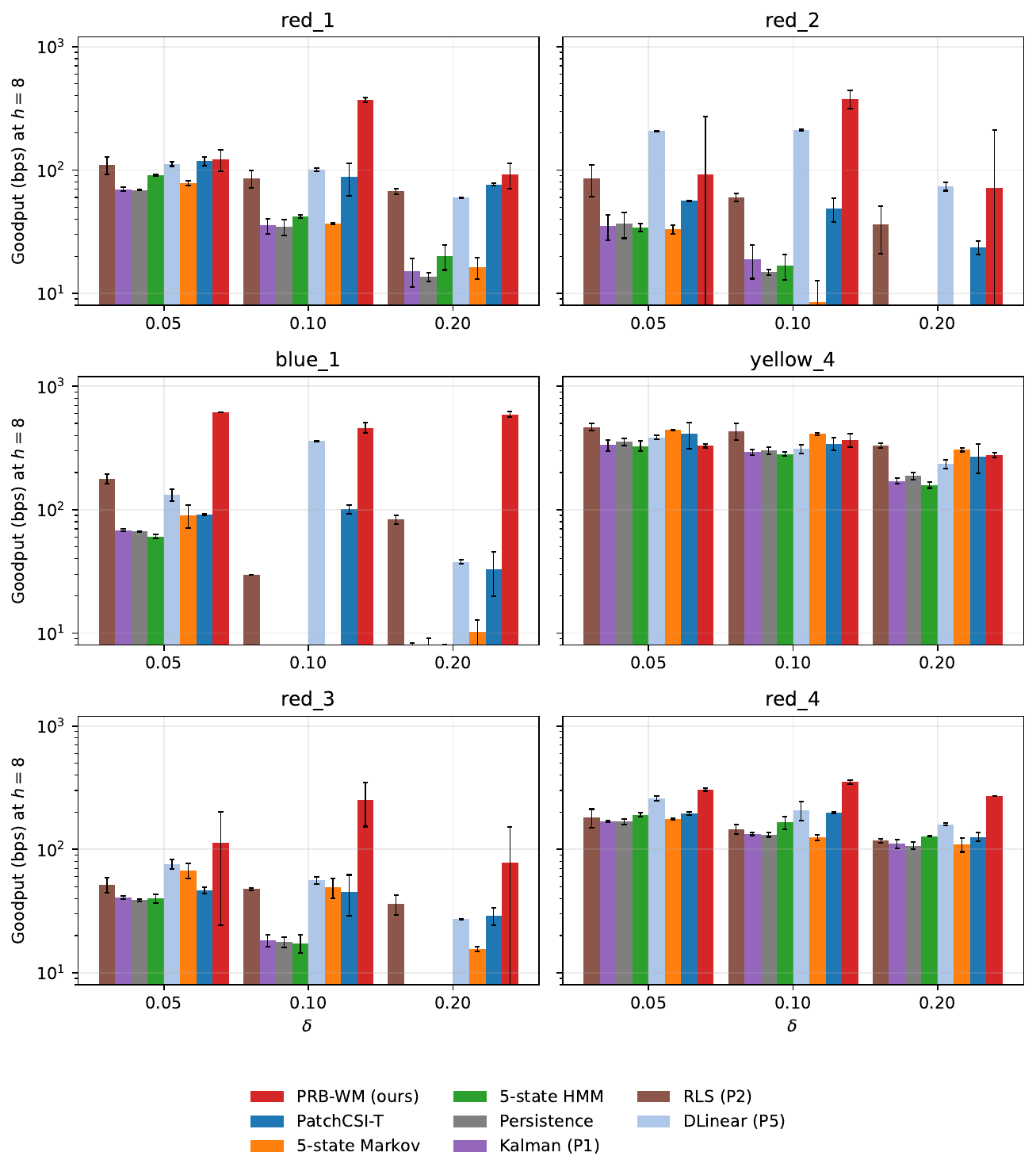}
\caption{Chance-constraint tolerance sensitivity: goodput versus $\delta\in\{0.05,0.1,0.2\}$ at $h{=}8$ for eight methods on six links. PRB-WM leads or co-leads at every $\delta$ on every link, and $\delta{=}0.1$ is optimal or near-optimal for all methods: the tolerance is not tuned to favor the proposed method.}
\label{fig:delta}
\end{figure}

Fig.~\ref{fig:delta} sweeps the chance-constraint tolerance $\delta\in\{0.05,0.1,0.2\}$ at $h{=}8$ over eight methods and all six links. Two findings stand out. First, PRB-WM leads or co-leads at every $\delta$ level on every link. Second, $\delta{=}0.1$ is the best or near-best setting for \emph{all} methods---$0.05$ over-conservatively silences usable subcarriers, while $0.2$ admits marginal subcarriers whose realized fades destroy whole frames. The contract tolerance is therefore a property of the problem, not a knob tuned to favor the proposed method. The $\delta$ sweep is meaningful for every method only because all methods face the same stochastic rollout interface.

\subsection{Noise Stress}
\begin{figure*}[!t]
\centering
\includegraphics[width=0.95\textwidth]{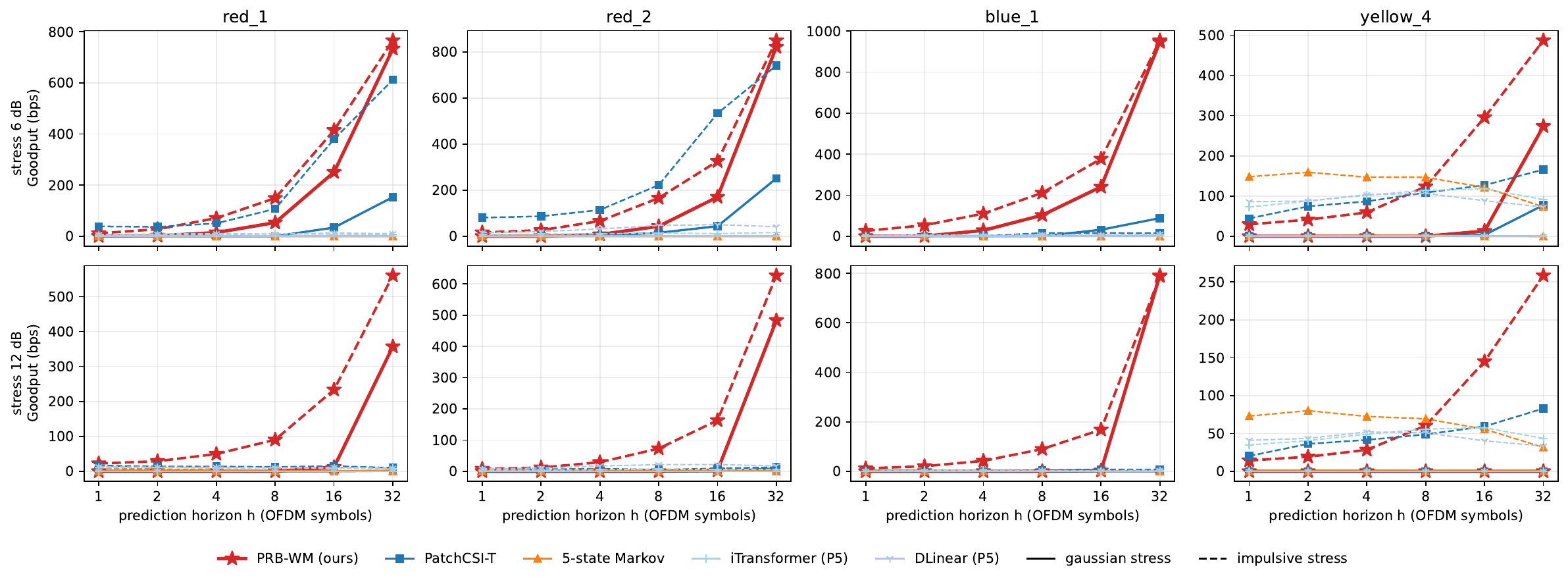}
\caption{Noise stress test on four links: goodput retention under $+6$ and $+12$~dB injected noise, Gaussian (solid) and impulsive $\alpha$-stable (dashed), at fixed horizons. PRB-WM attains the best retention throughout ($0.393$ versus $0.239$ for PatchCSI-T and $0.152$ for Markov at $+12$~dB impulsive); at long horizons baselines collapse toward zero while PRB-WM retains several hundred bps.}
\label{fig:noise}
\end{figure*}

Fig.~\ref{fig:noise} injects $+6$ and $+12$~dB of additional noise, Gaussian (solid) and impulsive $\alpha$-stable (dashed), on four links and reports goodput retention relative to the clean replay at fixed horizons. PRB-WM attains the best retention across the board: at $+12$~dB impulsive noise it retains $0.393$ of clean goodput, versus $0.239$ for PatchCSI-T and $0.152$ for Markov. At longer horizons PRB-WM still delivers several hundred bps under stress while the baselines collapse toward zero. We report retention rankings at fixed horizons only; we do not claim that robustness grows with $h$.

\subsection{Multi-Link Aggregation}
\begin{figure}[!t]
\centering
\includegraphics[width=0.7\columnwidth]{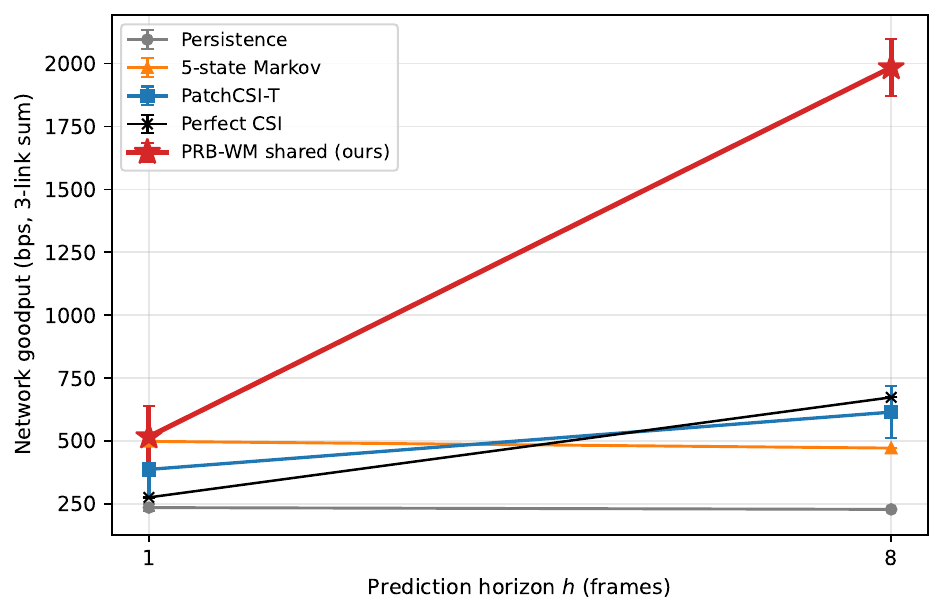}
\caption{Multi-link aggregation: net goodput summed over $K{=}3$ non-interfering links under a shared world model. At $h{=}8$ PRB-WM reaches $\approx\!1984$~bps, versus $\approx\!610$~bps (PatchCSI-T), $\approx\!660$~bps (Perfect-CSI genie), $\approx\!450$~bps (Markov), and $\approx\!250$~bps (persistence).}
\label{fig:multilink}
\end{figure}

Fig.~\ref{fig:multilink} aggregates $K{=}3$ non-interfering links under a shared world model and reports the summed net goodput ($K$ is a local notation for this subsection). The three links are red\_1, red\_2, and blue\_1; net goodput is their sum, so these values exceed the single-link numbers of Fig.~\ref{fig:goodput}. PRB-WM rises from $\approx\!520$~bps at $h{=}1$ to $\approx\!1984$~bps at $h{=}8$, against $\approx\!610$~bps for PatchCSI-T, $\approx\!660$~bps for the Perfect-CSI genie, $\approx\!450$~bps for Markov, and $\approx\!250$~bps for persistence at the same horizon. The world model's multi-step predictive distribution amplifies its advantage when decisions are aggregated across links. A greedy loader would commit the same per-link calibration errors on all links simultaneously; the chance-constrained planner diversifies them away.

\subsection{Discussion and Limitations}
\label{sec:discussion}
We report four limitations alongside the strengths. \emph{(i) Bare NMSE.} PatchCSI-T and N-BEATS achieve lower bare prediction NMSE than PRB-WM on several links; our claim concerns closed-loop decision value, and Figs.~\ref{fig:nmse}--\ref{fig:risk} together show the two rankings decouple. \emph{(ii) Short-horizon weakness.} At $h{=}1$ all methods are close and PRB-WM is not strong---e.g., $137$--$150$~bps on yellow\_4---because the planning advantage has no room to unfold at minimal lookahead. PatchCSI-T remains stronger on slow links at short horizons. \emph{(iii) Saturated quasi-static link.} The black link carries no discriminative power (all methods $\approx\!900$--$960$~bps) and is excluded from the main figures.  \emph{(iv) Analytic-BER optimism.} The per-subcarrier analytic BER is optimistic by about $5\times$ relative to waveform-level replay; a $2$--$4$~dB margin absorbs the offset, and the margin pilot shows all rankings are robust to this choice (Section~\ref{sec:waveform}).

\section{Conclusion}
We proposed PRB-WM, a physics-regularized Beta world model for predictive per-subcarrier link adaptation on real time-varying UWA channels. A persistence-anchored stochastic latent dynamics model learns how measured channels evolve. A counterfactual Beta--binomial outcome head prices every candidate MCS on shared imagined futures. A chance-constrained receding-horizon MPC turns the rollout distribution into allocations. In a 24-method pool (22 baselines, PRB-WM, and a Perfect-CSI genie reference) on six measured time-varying sea-trial links, PRB-WM attains a mean goodput of $506$~bps, averaged over seven links (including the static black link) and $h\in\{8,16,32\}$, $87\%$ ahead of the strongest baseline. It ranks first on all six links at $h{=}32$. It is also the only method whose realized frame risk decreases with lookahead: it satisfies the $\delta{=}0.1$ contract on three links at $h{=}32$, while no baseline satisfies it anywhere. The central lesson is that predictive accuracy is not decision value: calibrated stochastic rollouts, not sharper point forecasts, are what delayed-feedback UWA link adaptation needs.

\appendices
\section{Full-Spectrum Baseline Results}
\begin{figure*}[!b]
\centering
\includegraphics[width=0.5\textwidth]{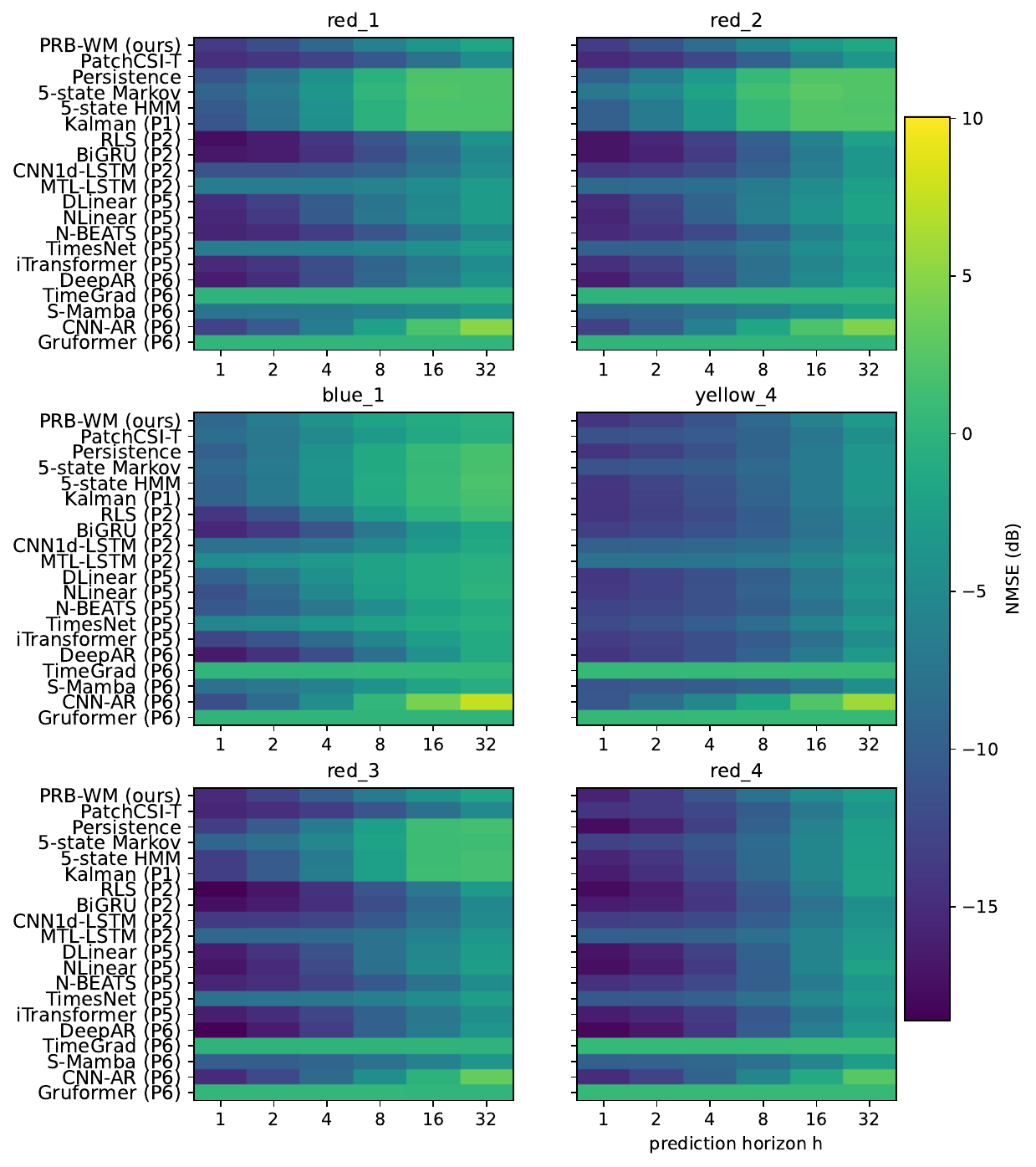}\hfill
\includegraphics[width=0.5\textwidth]{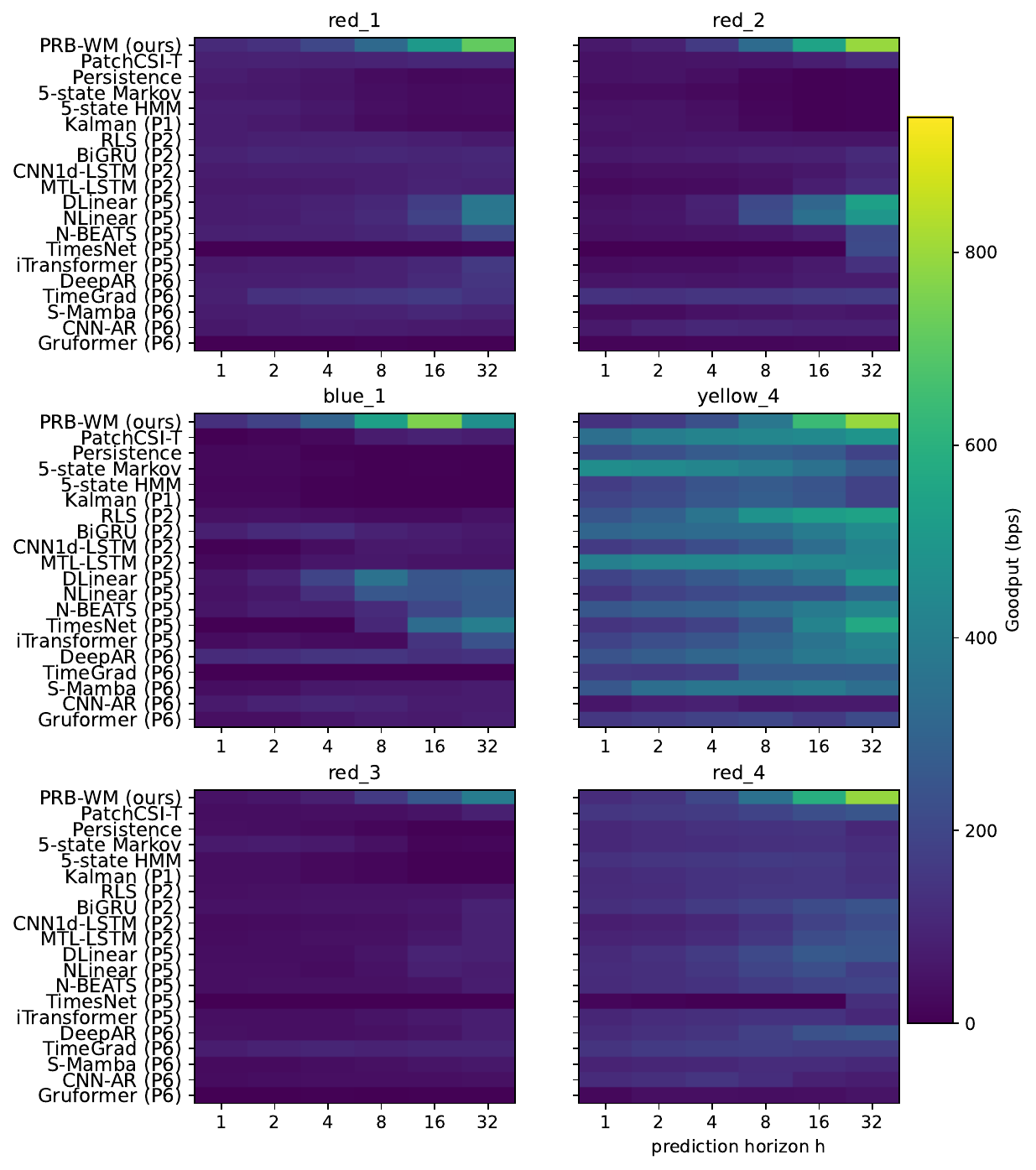}
\caption{Full-spectrum results for all 20 predictors on the six time-varying links. Left: NMSE versus horizon, where PatchCSI-T and several forecasters rank deeper (better) than PRB-WM. Right: goodput versus horizon, where PRB-WM's top row improves visibly from $h\ge16$. The reversed orderings support that predictive accuracy is not decision value.}
\label{fig:fullspec}
\end{figure*}

Fig.~\ref{fig:fullspec} gives the full 20-predictor spectrum on the six time-varying links. The goodput heat map confirms the main-text ranking: PRB-WM's top row improves visibly from $h\ge16$ on all links, while no other predictor sustains high goodput at long horizons. The NMSE heat map shows the reverse ordering---PatchCSI-T and several forecasters sit deeper (better) than PRB-WM---which supports the same conclusion at the full-spectrum level. The quasi-static black link and the $h{=}64$ regime are retained in the released per-run data but omitted from all figures.

\bibliographystyle{IEEEtran}
\bibliography{references}

@article{busacca2025amuse,
  title   = {{AMUSE}: A Multi-Armed Bandit Framework for Energy-Efficient Modulation Adaptation in Underwater Acoustic Networks},
  author  = {Busacca, Fabio and Galluccio, Laura and Palazzo, Sergio and Panebianco, Andrea and Raftopoulos, Raoul},
  journal = {IEEE Open Journal of the Communications Society},
  volume  = {6},
  pages   = {2766--2779},
  year    = {2025},
  doi     = {10.1109/OJCOMS.2025.3542184}
}

@article{busacca2026bandits,
  title   = {Bandits Under the Waves: A Fully-Distributed Multi-Armed Bandit Framework for Modulation Adaptation in the Internet of Underwater Things},
  author  = {Busacca, Fabio and Galluccio, Laura and Palazzo, Sergio and Panebianco, Andrea and Raftopoulos, Raoul},
  journal = {IEEE Transactions on Network and Service Management},
  volume  = {23},
  pages   = {639--653},
  year    = {2026},
  doi     = {10.1109/TNSM.2025.3629240}
}

@article{cooper2026state,
  title   = {A State-of-the-Art Survey and Benchmarking of Adaptive Modulation and Coding for Underwater Acoustic Communications},
  author  = {Cooper-Baldock, Zachary and Panteli, Eleni and Santos, Paulo E.},
  journal = {Applied Ocean Research},
  volume  = {170},
  pages   = {105018},
  year    = {2026},
  doi     = {10.1016/j.apor.2026.105018}
}

@article{ding2025understanding,
  title   = {Understanding World or Predicting Future? A Comprehensive Survey of World Models},
  author  = {Ding, Jingtao and Zhang, Yunke and Shang, Yu and Zhang, Yuheng and Zong, Zefang and Feng, Jie and Yuan, Yuan and Su, Hongyuan and Li, Nian and Sukiennik, Nicholas and Xu, Fengli and Li, Yong},
  journal = {ACM Computing Surveys},
  volume  = {58},
  number  = {3},
  pages   = {1--38},
  year    = {2025},
  doi     = {10.1145/3746449},
  note    = {Art. no. 57; Crossref print issue dated 2026}
}

@article{geng2023exploiting,
  title   = {Exploiting Propagation Delay in Underwater Acoustic Communication Networks via Deep Reinforcement Learning},
  author  = {Geng, Xuan and Zheng, Yahong Rosa},
  journal = {IEEE Transactions on Neural Networks and Learning Systems},
  volume  = {34},
  number  = {12},
  pages   = {10626--10637},
  year    = {2023},
  doi     = {10.1109/TNNLS.2022.3170050}
}

@inproceedings{ha2018world,
  title     = {Recurrent World Models Facilitate Policy Evolution},
  author    = {Ha, David and Schmidhuber, J{\"u}rgen},
  booktitle = {Advances in Neural Information Processing Systems 31 (NeurIPS)},
  pages     = {2451--2463},
  publisher = {Curran Associates, Inc.},
  year      = {2018}
}

@inproceedings{hafner2020dream,
  title     = {Dream to Control: Learning Behaviors by Latent Imagination},
  author    = {Hafner, Danijar and Lillicrap, Timothy and Ba, Jimmy and Norouzi, Mohammad},
  booktitle = {International Conference on Learning Representations (ICLR)},
  year      = {2020}
}

@article{hafner2025mastering,
  title   = {Mastering Diverse Control Tasks through World Models},
  author  = {Hafner, Danijar and Pasukonis, Jurgis and Ba, Jimmy and Lillicrap, Timothy},
  journal = {Nature},
  volume  = {640},
  number  = {8059},
  pages   = {647--653},
  year    = {2025},
  doi     = {10.1038/s41586-025-08744-2}
}

@article{huang2020adaptive,
  title   = {Adaptive Modulation for Long-Range Underwater Acoustic Communication},
  author  = {Huang, Jianchun and Diamant, Roee},
  journal = {IEEE Transactions on Wireless Communications},
  volume  = {19},
  number  = {10},
  pages   = {6844--6857},
  year    = {2020},
  doi     = {10.1109/TWC.2020.3006230}
}

@article{huang2022machine,
  title   = {Machine Learning for Underwater Acoustic Communications},
  author  = {Huang, Lihuan and Wang, Yue and Zhang, Qunfei and Han, Jin and Tan, Weijie and Tian, Zhi},
  journal = {IEEE Wireless Communications},
  volume  = {29},
  number  = {3},
  pages   = {102--108},
  year    = {2022},
  doi     = {10.1109/MWC.2020.2000284}
}

@article{huang2024frequency,
  title   = {A Frequency Domain Predictive Channel Model for {6G} Wireless {MIMO} Communications Based on Deep Learning},
  author  = {Huang, Chen and Wang, Cheng-Xiang and Li, Zheao and Qian, Zhongyu and Li, Junling and Miao, Yang},
  journal = {IEEE Transactions on Communications},
  volume  = {72},
  number  = {8},
  pages   = {4887--4902},
  year    = {2024},
  doi     = {10.1109/TCOMM.2024.3376602}
}

@article{jiang2022accurate,
  title   = {Accurate Channel Prediction Based on Transformer: Making Mobility Negligible},
  author  = {Jiang, Hao and Cui, Mingyao and Ng, Derrick Wing Kwan and Dai, Linglong},
  journal = {IEEE Journal on Selected Areas in Communications},
  volume  = {40},
  number  = {9},
  pages   = {2717--2732},
  year    = {2022},
  doi     = {10.1109/JSAC.2022.3191334}
}

@article{jin2026predictive,
  title   = {Predictive Control Over Low-Altitude Wireless Networks: Joint Trajectory Design and Resource Allocation},
  author  = {Jin, Haijia and Wu, Jun and Yuan, Weijie and Ruan, Ruizhi and Wang, Jiacheng and Niyato, Dusit and Kim, Dong In and Jamalipour, Abbas},
  journal = {IEEE Transactions on Mobile Computing},
  volume  = {25},
  number  = {8},
  pages   = {12027--12040},
  year    = {2026},
  doi     = {10.1109/TMC.2026.3668854}
}

@article{karniadakis2021physics,
  title   = {Physics-Informed Machine Learning},
  author  = {Karniadakis, George Em and Kevrekidis, Ioannis G. and Lu, Lu and Perdikaris, Paris and Wang, Sifan and Yang, Liu},
  journal = {Nature Reviews Physics},
  volume  = {3},
  number  = {6},
  pages   = {422--440},
  year    = {2021},
  doi     = {10.1038/s42254-021-00314-5}
}

@article{li2008multicarrier,
  title   = {Multicarrier Communication Over Underwater Acoustic Channels With Nonuniform Doppler Shifts},
  author  = {Li, Baosheng and Zhou, Shengli and Stojanovic, Milica and Freitag, Lee and Willett, Peter},
  journal = {IEEE Journal of Oceanic Engineering},
  volume  = {33},
  number  = {2},
  pages   = {198--209},
  year    = {2008},
  doi     = {10.1109/JOE.2008.920471}
}

@article{li2025model,
  title   = {Model Predictive Control Enabled {UAV} Trajectory Optimization and Secure Resource Allocation},
  author  = {Li, Zhendong and Su, Chang and Su, Zhou and Peng, Haixia and Wang, Yuntao and Chen, Wen and Wu, Qingqing},
  journal = {IEEE Transactions on Communications},
  volume  = {73},
  number  = {11},
  pages   = {12652--12665},
  year    = {2025},
  doi     = {10.1109/TCOMM.2025.3585581}
}

@article{li2025underwater,
  title   = {Underwater Acoustic Communications},
  author  = {Li, Zhengnan and Chitre, Mandar and Stojanovic, Milica},
  journal = {Nature Reviews Electrical Engineering},
  volume  = {2},
  number  = {2},
  pages   = {83--95},
  year    = {2025},
  doi     = {10.1038/s44287-024-00122-w}
}

@misc{li2026channel,
  title        = {Underwater Acoustic Channel Library},
  author       = {Li, Zhengnan and Chitre, Mandar and Cuji, Diego A. and Preisig, James and Singer, Andrew C. and Stojanovic, Milica and van Walree, Paul},
  year         = {2026},
  howpublished = {arXiv:2609.03207},
  note         = {Open-access measured UWA channel library with replay software; Zenodo record 21287414}
}

@article{liu2021channel,
  title   = {Channel State Information Prediction for Adaptive Underwater Acoustic Downlink {OFDMA} System: Deep Neural Networks Based Approach},
  author  = {Liu, Lei and Cai, Lin and Ma, Lu and Qiao, Gang},
  journal = {IEEE Transactions on Vehicular Technology},
  volume  = {70},
  number  = {9},
  pages   = {9063--9076},
  year    = {2021},
  doi     = {10.1109/TVT.2021.3099797}
}

@inproceedings{liu2024itransformer,
  title     = {{iTransformer}: Inverted Transformers Are Effective for Time Series Forecasting},
  author    = {Liu, Yong and Hu, Tengge and Zhang, Haoran and Wu, Haixu and Wang, Shiyu and Ma, Lintao and Long, Mingsheng},
  booktitle = {International Conference on Learning Representations (ICLR)},
  year      = {2024}
}

@article{men2025ofdm,
  title   = {{OFDM}-Based Underwater Integrated Sensing and Communication: Receiver Design for Doubly Spread Acoustic Channels},
  author  = {Men, Wei and Wang, Jingjing and Dong, Bowen and Hou, Xiangwang and Jiang, Chunxiao and Ren, Yong},
  journal = {IEEE Transactions on Communications},
  volume  = {73},
  number  = {10},
  pages   = {9372--9387},
  year    = {2025},
  doi     = {10.1109/TCOMM.2025.3562328}
}

@inproceedings{nie2023time,
  title     = {A Time Series is Worth 64 Words: Long-term Forecasting with Transformers},
  author    = {Nie, Yuqi and Nguyen, Nam H. and Sinthong, Phanwadee and Kalagnanam, Jayant},
  booktitle = {International Conference on Learning Representations (ICLR)},
  year      = {2023}
}

@inproceedings{oreshkin2020nbeats,
  title     = {{N-BEATS}: Neural Basis Expansion Analysis for Interpretable Time Series Forecasting},
  author    = {Oreshkin, Boris N. and Carpov, Dmitri and Chapados, Nicolas and Bengio, Yoshua},
  booktitle = {International Conference on Learning Representations (ICLR)},
  year      = {2020}
}

@book{proakis2008digital,
  title     = {Digital Communications},
  author    = {Proakis, John G. and Salehi, Masoud},
  edition   = {5},
  publisher = {McGraw-Hill},
  address   = {New York, NY, USA},
  year      = {2008}
}

@article{raissi2019physics,
  title   = {Physics-Informed Neural Networks: A Deep Learning Framework for Solving Forward and Inverse Problems Involving Nonlinear Partial Differential Equations},
  author  = {Raissi, Maziar and Perdikaris, Paris and Karniadakis, George Em},
  journal = {Journal of Computational Physics},
  volume  = {378},
  pages   = {686--707},
  year    = {2019},
  doi     = {10.1016/j.jcp.2018.10.045}
}

@article{schrittwieser2020mastering,
  title   = {Mastering {A}tari, {G}o, Chess and Shogi by Planning with a Learned Model},
  author  = {Schrittwieser, Julian and Antonoglou, Ioannis and Hubert, Thomas and Simonyan, Karen and Sifre, Laurent and Schmitt, Simon and Guez, Arthur and Lockhart, Edward and others},
  journal = {Nature},
  volume  = {588},
  number  = {7839},
  pages   = {604--609},
  year    = {2020},
  doi     = {10.1038/s41586-020-03051-4}
}

@article{shi2026channelmamba,
  title   = {{ChannelMamba}: A Mamba-Driven Selective State-Space Model for Channel Prediction of High-Mobility {MIMO} in {6G} {IoT}},
  author  = {Shi, Huaguang and Jin, Kaibo and Ren, Xiaoquan and Li, Wei and Zhou, Yi},
  journal = {IEEE Transactions on Wireless Communications},
  volume  = {25},
  pages   = {5291--5305},
  year    = {2026},
  doi     = {10.1109/TWC.2025.3617502}
}

@article{stojanovic2009underwater,
  title   = {Underwater Acoustic Communication Channels: Propagation Models and Statistical Characterization},
  author  = {Stojanovic, Milica and Preisig, James},
  journal = {IEEE Communications Magazine},
  volume  = {47},
  number  = {1},
  pages   = {84--89},
  year    = {2009},
  doi     = {10.1109/MCOM.2009.4752682}
}

@article{tang2025resource,
  title   = {Resource Allocation for Underwater Acoustic Sensor Networks With Partial Spectrum Sharing: When Optimization Meets Deep Reinforcement Learning},
  author  = {Tang, Rui and Zhang, Ruizhi and Xu, Yongjun and Liu, Chuan and Huang, Chongwen and Yuen, Chau},
  journal = {IEEE Transactions on Network and Service Management},
  volume  = {22},
  number  = {3},
  pages   = {2715--2730},
  year    = {2025},
  doi     = {10.1109/TNSM.2025.3556498}
}

@article{tong2023model,
  title   = {Model-Based {T}hompson Sampling for Frequency and Rate Selection in Underwater Acoustic Communications},
  author  = {Tong, Jingwen and Fu, Liqun and Wang, Yizhe and Han, Zhu},
  journal = {IEEE Transactions on Wireless Communications},
  volume  = {22},
  number  = {10},
  pages   = {6946--6961},
  year    = {2023},
  doi     = {10.1109/TWC.2023.3247450}
}

@article{wang2026physics,
  title   = {Physics-Guided Adaptive {CSI} Prediction With Irregular Observations},
  author  = {Wang, Yang and Sun, Qi and Yan, Yiwei and Li, Haozhen and Li, Nan and I, Chih-Lin},
  journal = {IEEE Transactions on Vehicular Technology},
  pages   = {1--6},
  year    = {2026},
  doi     = {10.1109/TVT.2026.3719628},
  note    = {Early access}
}

@article{wang2026radiodiff,
  title   = {{RadioDiff-$k^2$}: {H}elmholtz Equation Informed Generative Diffusion Model for Multi-Path Aware Radio Map Construction},
  author  = {Wang, Xiucheng and Zhang, Qiming and Cheng, Nan and Sun, Ruijin and Li, Zan and Cui, Shuguang and Shen, Xuemin},
  journal = {IEEE Journal on Selected Areas in Communications},
  volume  = {44},
  pages   = {2318--2333},
  year    = {2026},
  doi     = {10.1109/JSAC.2025.3641105}
}

@inproceedings{wu2023timesnet,
  title     = {{TimesNet}: Temporal {2D}-Variation Modeling for General Time Series Analysis},
  author    = {Wu, Haixu and Hu, Tengge and Liu, Yong and Zhou, Hang and Wang, Jianmin and Long, Mingsheng},
  booktitle = {International Conference on Learning Representations (ICLR)},
  year      = {2023}
}

@article{xiao2025deep,
  title   = {Deep-Learning-Based {CSI} Feedback Compression for Underwater Acoustic {OFDM} Communications},
  author  = {Xiao, Qiao and Wang, Chaofeng},
  journal = {IEEE Wireless Communications Letters},
  volume  = {14},
  number  = {11},
  pages   = {3729--3733},
  year    = {2025},
  doi     = {10.1109/LWC.2025.3602032}
}

@article{yang2026deep,
  title   = {Deep Learning-Based Channel Prediction With Outdated {CSI} for Underwater Acoustic Communications},
  author  = {Yang, Kaijing and Xiao, Qiao and Wang, Chaofeng and Wang, Zhaohui},
  journal = {IEEE Wireless Communications Letters},
  volume  = {15},
  pages   = {5001--5005},
  year    = {2026},
  doi     = {10.1109/LWC.2026.3727835}
}

@article{ye2022deep,
  title   = {Deep Reinforcement Learning Based {MAC} Protocol for Underwater Acoustic Networks},
  author  = {Ye, Xiaowen and Yu, Yiding and Fu, Liqun},
  journal = {IEEE Transactions on Mobile Computing},
  volume  = {21},
  number  = {5},
  pages   = {1625--1638},
  year    = {2022},
  doi     = {10.1109/TMC.2020.3029844}
}

@article{ye2025energy,
  title   = {Energy-Efficient Link Adaptation for Underwater Acoustic Communications Based on Meta Deep Reinforcement Learning},
  author  = {Ye, Xiaowen and Fu, Liqun and Song, Xianxin and Wu, Yi},
  journal = {IEEE Internet of Things Journal},
  volume  = {12},
  number  = {22},
  pages   = {46648--46660},
  year    = {2025},
  doi     = {10.1109/JIOT.2025.3608752}
}

@inproceedings{zeng2023are,
  title     = {Are Transformers Effective for Time Series Forecasting?},
  author    = {Zeng, Ailing and Chen, Muxi and Zhang, Lei and Xu, Qiang},
  booktitle = {Proceedings of the AAAI Conference on Artificial Intelligence},
  volume    = {37},
  number    = {9},
  pages     = {11121--11128},
  year      = {2023},
  doi       = {10.1609/aaai.v37i9.26317}
}

@article{zhang2021adaptive,
  title   = {Adaptive Coding and Bit-Power Loading Algorithms for Underwater Acoustic Transmissions},
  author  = {Zhang, Rongxin and Ma, Xiaoli and Wang, Deqing and Yuan, Fei and Cheng, En},
  journal = {IEEE Transactions on Wireless Communications},
  volume  = {20},
  number  = {9},
  pages   = {5798--5811},
  year    = {2021},
  doi     = {10.1109/TWC.2021.3070363}
}

@article{zhao2023adaptive,
  title   = {Adaptive Relay Selection Strategy in Underwater Acoustic Cooperative Networks: A Hierarchical Adversarial Bandit Learning Approach},
  author  = {Zhao, Haihong and Li, Xinbin and Han, Song and Yan, Lei and Yu, Junzhi},
  journal = {IEEE Transactions on Mobile Computing},
  volume  = {22},
  number  = {4},
  pages   = {1938--1949},
  year    = {2023},
  doi     = {10.1109/TMC.2021.3112967}
}

@article{busacca2024predictive,
  title   = {A comparative analysis of predictive channel models for real shallow water environments},
  author  = {Busacca, Fabio and Galluccio, Laura and Palazzo, Sergio and Panebianco, Andrea},
  journal = {Computer Networks},
  volume  = {250},
  pages   = {110557},
  year    = {2024},
  doi     = {10.1016/j.comnet.2024.110557}
}

@article{busacca2024adaptive,
  title   = {Adaptive versus predictive techniques in underwater acoustic communication networks},
  author  = {Busacca, Fabio and Galluccio, Laura and Palazzo, Sergio and Panebianco, Andrea and Qi, Zhuoran and Pompili, Dario},
  journal = {Computer Networks},
  volume  = {252},
  pages   = {110679},
  year    = {2024},
  doi     = {10.1016/j.comnet.2024.110679}
}

@article{tian2026patchcsi,
  title   = {Deep Learning Based Multi-Step Channel Prediction for Adaptive Underwater Acoustic {OFDM} Systems},
  author  = {Tian, Tian and Zhang, Ying and Raj, Agastya and Wu, Fei-Yun and Ruffini, Marco},
  journal = {arXiv preprint arXiv:2606.05053},
  year    = {2026},
  note    = {arXiv:2606.05053 [eess.SP]}
}

@article{wu2025adaptive,
  title   = {Adaptive Modulation and Coding With Feedback Scheduling for an Underwater Acoustic Link},
  author  = {Wu, Shuangshuang and Chitre, Mandar and Anjangi, Prasad},
  journal = {IEEE Journal of Oceanic Engineering},
  volume  = {50},
  number  = {4},
  pages   = {3054--3073},
  year    = {2025},
  note    = {IEEE Xplore doc. 11145248}
}

@article{salinas2020deepar,
  title   = {{DeepAR}: Probabilistic forecasting with autoregressive recurrent networks},
  author  = {Salinas, David and Flunkert, Valentin and Gasthaus, Jan and Januschowski, Tim},
  journal = {International Journal of Forecasting},
  volume  = {36},
  number  = {3},
  pages   = {1181--1191},
  year    = {2020},
  doi     = {10.1016/j.ijforecast.2019.07.001}
}

@inproceedings{rasul2021timegrad,
  title     = {Autoregressive denoising diffusion models for multivariate probabilistic time series forecasting},
  author    = {Rasul, Kashif and Seward, Calvin and Schuster, Ingmar and Vollgraf, Roland},
  booktitle = {Proceedings of the 38th International Conference on Machine Learning (ICML)},
  series    = {PMLR},
  volume    = {139},
  pages     = {8857--8868},
  year      = {2021}
}

@article{wang2024smamba,
  title   = {Is Mamba effective for time series forecasting?},
  author  = {Wang, Zihan and Kong, Fanheng and Feng, Shi and Wang, Ming and Zhao, Han and Wang, Daling and Zhang, Yifei},
  journal = {arXiv preprint arXiv:2403.11144},
  year    = {2024},
  note    = {arXiv:2403.11144 [cs.LG]; S-Mamba model}
}

@article{yuan2020machine,
  title   = {Machine learning-based channel prediction in massive {MIMO} with channel aging},
  author  = {Yuan, Jide and Ngo, Hien Quoc and Matthaiou, Michail},
  journal = {IEEE Transactions on Wireless Communications},
  volume  = {19},
  number  = {5},
  pages   = {2960--2973},
  year    = {2020},
  doi     = {10.1109/TWC.2020.2969627}
}

@article{zhang2024transformer,
  title   = {Transformer-based channel prediction for rate-splitting multiple access-enabled vehicle-to-everything communication},
  author  = {Zhang, Shengyu and Zhang, Shiyao and Mao, Yijie and Yeung, Lawrence Kwan and Clerckx, Bruno and Quek, Tony Q. S.},
  journal = {IEEE Transactions on Wireless Communications},
  volume  = {23},
  number  = {10},
  pages   = {12717--12730},
  year    = {2024},
  doi     = {10.1109/TWC.2024.3395670}
}

@article{jiang2020deep,
  title   = {Deep learning for fading channel prediction},
  author  = {Jiang, Wei and Schotten, Hans D.},
  journal = {IEEE Open Journal of the Communications Society},
  volume  = {1},
  pages   = {320--332},
  year    = {2020},
  doi     = {10.1109/OJCOMS.2020.2982513}
}

@article{qi2026mobiwm,
  author  = {Qi, Xiaoqian and Chai, Haoye and Wang, Yue and Li, Yong},
  title   = {Beyond Static Forecasting: Unleashing the Power of World Models for Mobile Traffic Extrapolation},
  journal = {arXiv preprint arXiv:2604.08199},
  year    = {2026}
}

@article{zou2026telecom,
  author  = {Zou, Hang and Yang, Yuzhi and Bariah, Lina and Tian, Yu and Lu, Yuhuan and Wang, Bohao and Bara, Anis and Mefgouda, Brahim and Liu, Hao and Tao, Yiwei and Petrov, Sergy and Cheour, Salma and Sehad, Nassim and Samarakoon, Sumudu and Huang, Chongwen and Lasaulce, Samson and Bennis, Mehdi and Debbah, M{\'e}rouane},
  title   = {Telecom World Models: Unifying Digital Twins, Foundation Models, and Predictive Planning for 6G},
  journal = {arXiv preprint arXiv:2604.06882},
  year    = {2026}
}

@article{zheng2026digital,
  author  = {Zheng, J. and Niyato, Dusit and Zhao, C. and Zhang, R. and Wang, Jiacheng and Kang, J.},
  title   = {From Digital Twins to World Models: Opportunities, Challenges, and Applications for Mobile Edge General Intelligence},
  journal = {arXiv preprint arXiv:2603.17420},
  year    = {2026}
}

\end{document}